\documentclass[twocolumn,showpacs,preprintnumbers,superscriptaddress,amsfonts,prd,floatfix,tightenlines]{revtex4}
\usepackage{graphicx}% Include figure files
\usepackage{dcolumn}% Align table columns on decimal point
\usepackage{longtable}%
\usepackage{multirow}
\usepackage[dvips]{epsfig}
\usepackage{bm}% bold math

\def\PRETH{{\sc pretrigger hi}}
\def\GLL{{\sc local$\otimes$global lo}}
\def\SLL{{\sc single local lo}}
\def\SLH{{\sc single local hi}}

\def\LOCALH{{\sc local hi}}
\def\LOCALL{{\sc local lo}}
\def\GLOBALL{{\sc global lo}}

\newcommand{\ycm}{\mbox{$y_{\rm{cm}}$}}

\def\DIFFXS{Ed \sigma/d^{3}p}

\def\pCu{${\mit p}$Cu}

\def\piCu{${\pi^{-}}$Cu}

\def\UCDAVIS{University of California-Davis, Davis, California 95616}
\def\MSU{Michigan State University, East Lansing, Michigan 48824}
\def\Delhi{University of Delhi, Delhi, India 110007}
\def\FNAL{Fermi National Accelerator Laboratory, Batavia,
                   Illinois 60510}
\def\NEU{Northeastern University, Boston, Massachusetts  02115}
\def\OK{University of Oklahoma, Norman, Oklahoma  73019}
\def\PSU{Pennsylvania State University, University Park,
		   Pennsylvania 16802}
\def\PU{University of Pittsburgh, Pittsburgh, Pennsylvania 15260}
\def\UR{University of Rochester, Rochester, New York 14627}

\def\bbcoord{20 150 652 652}
\def\bbcoordb{20 400 652 652}

\def\figsize{4.2in}

\begin{document}
%\preprint{FERMILAB-Pub-05/036-E}
\title{ Nuclear effects in high-$\bm p_T$ production of 
direct photons and neutral mesons}

\date{\today}

\author{L.~Apanasevich}\affiliation{\MSU}\affiliation{\UR}
\author{J.~Bacigalupi}\affiliation{\UCDAVIS}
\author{W.~Baker}\affiliation{\FNAL}
\author{M.~Begel}\affiliation{\UR}
\author{S.~Blusk}\affiliation{\PU}
\author{C.~Bromberg}\affiliation{\MSU}
\author{P.~Chang}\affiliation{\NEU}
\author{B.~Choudhary}\affiliation{\Delhi}
\author{W.~H.~Chung}\affiliation{\PU}
\author{L.~de~Barbaro}\affiliation{\UR}
\author{W.~DeSoi}\affiliation{\UR}
\author{W.~D{\l}ugosz}\affiliation{\NEU}
\author{J.~Dunlea}\affiliation{\UR}
\author{E.~Engels,~Jr.}\affiliation{\PU}
\author{G.~Fanourakis}\affiliation{\UR}
\author{T.~Ferbel}\affiliation{\UR}
\author{J.~Ftacnik}\affiliation{\UR}
\author{D.~Garelick}\affiliation{\NEU}
\author{G.~Ginther}\affiliation{\UR}
\author{M.~Glaubman}\affiliation{\NEU}
\author{P.~Gutierrez}\affiliation{\OK}
\author{K.~Hartman}\affiliation{\PSU}
\author{J.~Huston}\affiliation{\MSU}
\author{C.~Johnstone}\affiliation{\FNAL}
\author{V.~Kapoor}\affiliation{\Delhi}
\author{J.~Kuehler}\affiliation{\OK}
\author{C.~Lirakis}\affiliation{\NEU}
\author{F.~Lobkowicz}\altaffiliation{Deceased}\affiliation{\UR}
\author{P.~Lukens}\affiliation{\FNAL}
\author{J.~Mansour}\affiliation{\UR}
\author{A.~Maul}\affiliation{\MSU}
\author{R.~Miller}\affiliation{\MSU}
\author{B.~Y.~Oh}\affiliation{\PSU}
\author{G.~Osborne}\affiliation{\UR}
\author{D.~Pellett}\affiliation{\UCDAVIS}
\author{E.~Prebys}\affiliation{\UR}
\author{R.~Roser}\affiliation{\UR}
\author{P.~Shepard}\affiliation{\PU}
\author{R.~Shivpuri}\affiliation{\Delhi}
\author{D.~Skow}\affiliation{\FNAL}
\author{P.~Slattery}\affiliation{\UR}
\author{L.~Sorrell}\affiliation{\MSU}
\author{D.~Striley}\affiliation{\NEU}
\author{W.~Toothacker}\altaffiliation{Deceased}\affiliation{\PSU}
\author{S.~M.~Tripathi}\affiliation{\UCDAVIS}
\author{N.~Varelas}\affiliation{\UR}
\author{D.~Weerasundara}\affiliation{\PU}
\author{J.~J.~Whitmore}\affiliation{\PSU}
\author{T.~Yasuda}\affiliation{\NEU}
\author{C.~Yosef}\affiliation{\MSU}
\author{M.~Zieli\'{n}ski}\affiliation{\UR}
\author{V.~Zutshi}\affiliation{\Delhi}
\collaboration{Fermilab E706 Collaboration}\noaffiliation

\begin{abstract}
We present results on the production of direct photons, $\pi^0$, and
$\eta$ mesons on nuclear targets at large transverse momenta ($p_T$).  
The data are from 530 and 800~GeV/c proton beams and 515~GeV/c $\pi^-$ beams
incident upon copper and beryllium targets that span the
kinematic range of $1.0 < p_T \alt 10$~GeV/c at central rapidities.
\end{abstract}

\maketitle

\section{Introduction}

The study of inclusive particle production at large transverse momenta
($p_T$) has yielded valuable information about perturbative quantum
chromodynamics (PQCD), parton distribution functions (PDF), and
fragmentation functions of
partons~\cite{geist,mccubbin,owens,molzon,kt-dp}.  The use of
nuclear targets provides, in addition, information on parton and hadron
rescattering and explores the time evolution of the collision.  Since 
the discovery of the nuclear enhancement of high-$p_T$ single-particle
production~\cite{cronin,antreasyan,frish}, a large
body of data has been accumulated to investigate nuclear-target
effects in a wide variety of production processes, including those
yielding single hadrons, dihadron pairs, Drell-Yan pairs, two-jet
systems, and heavy flavors.  Recent results from the RHIC
program~\cite{star,phenix,phobos,brahms}, in particular, have
highlighted the differences between initial and final-state effects
in the nuclear environment.  Many approaches have been developed to 
explain these data, which have included models
for multiple-scattering, Fermi motion,
modification of parton densities in the nuclear medium, QCD
higher-twist contributions, and new states of matter.

We present the results of a high-statistics study of
nuclear effects in the inclusive production of direct photons,
$\pi^0$ and $\eta$ mesons at large $p_T$ using data from Fermilab
experiment E706, and compare the results to predictions of a phenomenological 
model of nuclear effects~\cite{wang-note}.

\section{Apparatus}

\subsection{Meson West spectrometer}

Fermilab E706 was a fixed-target experiment designed to measure the
production of direct photons, neutral mesons, and associated particles
at
high-$p_T$~\cite{E706-kt,E706-pos-pieta,E706-neg-pieta,E706-directphoton,E706-charm,E706-omega}.
The apparatus included a charged particle spectrometer and a large
liquid argon calorimeter, as described below.  Additional information
about the Meson West spectrometer can be found in earlier
papers~\cite{E706-charm,E706-calibration}.

\begin{figure}
\vskip-1.75truein
\epsfxsize=3.8truein
\epsfbox[20 300 652 652]{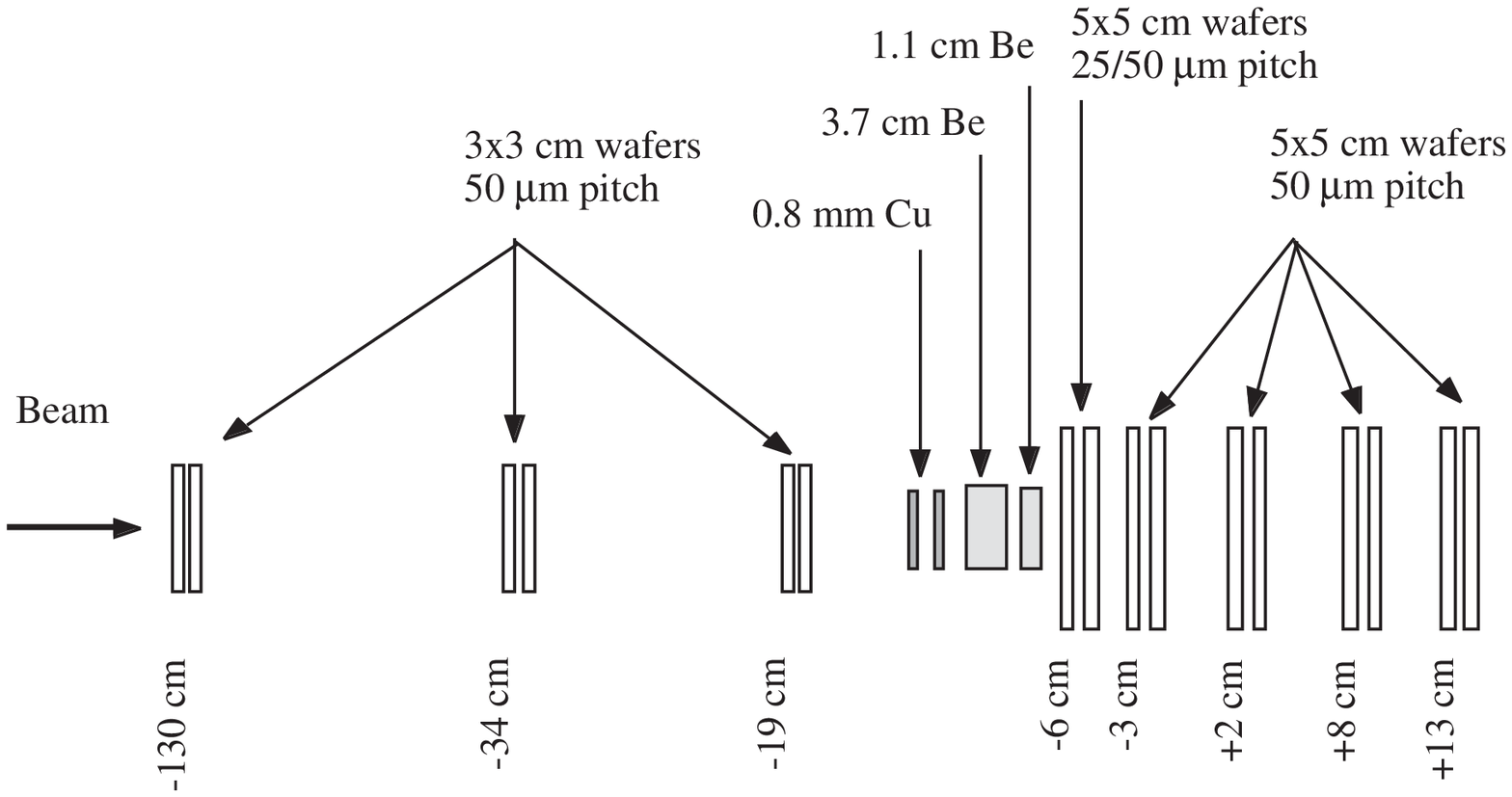}
\vskip-0.5truecm
\epsfxsize=3.8truein
\epsfbox[20 300 652 652]{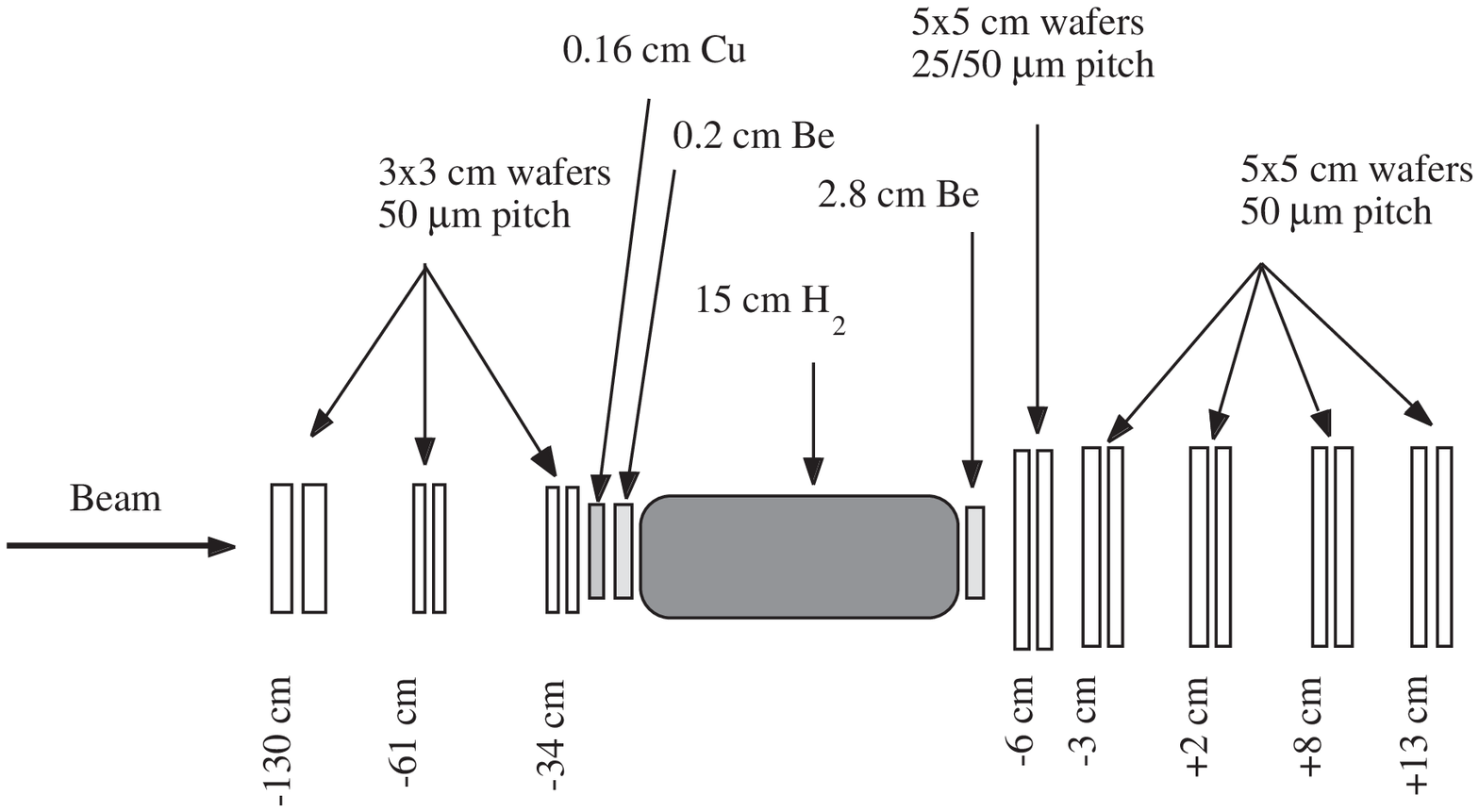}
\vskip4truecm
\caption{Target configuration during the 1990 (top) and 1991--1992 (bottom)
fixed target runs.}
\label{fig:target}
\end{figure}

This paper reports on data from the two primary data runs of the
experiment.  During the 1990 run, the target consisted of two 0.8~mm
thick copper foils followed by two pieces of
beryllium~(Fig.~\ref{fig:target}:top).  The upstream piece of
beryllium was 3.7~cm long, while the length of the downstream piece
was 1.1~cm.  In the 1991--1992 run, the target consisted of two 0.8~mm
thick copper foils immediately upstream of a liquid hydrogen
target~\cite{h2target}, followed by a 2.54~cm long beryllium
cylinder~(Fig.~\ref{fig:target}:bottom).  The liquid hydrogen was
contained in a 15.3~cm long mylar flask, which was supported in an
evacuated volume with beryllium windows at each end (2.5~mm thickness
upstream and 2.8~mm thickness downstream).  The target
material is detailed in Table~\ref{tab:target}.
%%%%%%%%%%%%%%
%%%%%%%%%%%%
%%%%%%%%%%%%

%  Target    Length   Density    LamdaI   Absorption Length
%
%  530 GeV/c p Beam
%    BE      3.070    1.855      77.736      0.073259      
%    CU      0.156    8.810     134.888      0.010189
%    SI      0.180    2.330     107.094      0.003916
%    H2     14.500    0.071      50.369      0.020439
%
%  800 GeV/c p Beam
%    BE      3.070    1.855      76.181      0.074754     
%    CU      0.156    8.810     132.190      0.010397
%    SI      0.180    2.330     104.952      0.003996
%    H2     14.500    0.071      49.362      0.020856
%
%  515 GeV/c pi- Beam (1991)
%    BE      3.070    1.855     104.082     0.054715
%    CU      0.156    8.810     167.930     0.0081841
%    SI      0.180    2.330     137.512     0.0030499
%    H2     14.500    0.071      76.910     0.013386
%
%  515 GeV/c pi- Beam (1990)
%    BE      4.829    1.848     104.082     0.085740 
%    CU      0.156    8.810     167.930     0.0081841
%    SI      0.240    2.330     137.512     0.0040666
\begin{table}[h]
\caption{Target fiducial lengths, densities, and number of interaction lengths
by sample. The interaction lengths for the nuclear targets were obtained 
from Ref.~\cite{carroll}; for the H$_2$ target they were obtained from 
Ref.~\cite{Baldini:1988ti}.}
\begin{tabular}{clcccc}
Target        & \multicolumn{1}{c}{Beam}   & Target   & Fiducial & Measured     &  Interaction \\
Layout        & \multicolumn{1}{c}{(GeV/$c$)}&        & Length   & Density      &   Length     \\
              &                            &          &  (cm)    & (g/cm${}^3$) & (g/cm${}^2$) \\
\hline\hline
\multirow{2}{*}{1990}
              & \multirow{2}{*}[.3ex]{515 $\pi^-$} & Be       & 4.829       & 1.848       & 0.086 \\
              &             & Cu       & 0.156       & 8.810       & 0.008 \\
\hline
              & \ 515 $\pi^-$ &          &             &             & 0.055 \\
              & \ 530 $p$     & Be       & 3.070       & 1.855       & 0.073 \\
              & \ 800 $p$     &          &             &             & 0.074 \\
\cline{2-6}
              & \ 515 $\pi^-$ &          &             &             & 0.008 \\
1991--1992    & \ 530 $p$     & Cu       & 0.156       & 8.810       & 0.010 \\
              & \ 800 $p$     &          &             &             & 0.010 \\
\cline{2-6}
              & \ 515 $\pi^-$ &          &             &             & 0.004 \\
              & \ 530 $p$     & H{${}_2$}& 14.500      & 0.0705      & 0.020 \\
              & \ 800 $p$     &          &             &             & 0.021 \\
\hline\hline
\end{tabular}
\label{tab:target}
\end{table}
%%%%%%%%%%%%
 
The charged particle spectrometer consisted of silicon microstrip
detectors (SSDs) in the target region and multiwire proportional
chambers (PWCs) and straw tube drift chambers (STDCs) downstream of a
large-aperture analysis magnet~\cite{E706-charm}.  Six
3$\times$3~cm$^2$ SSD planes were located upstream of the target
region and used to reconstruct beam tracks.  Two hybrid 5$\times
$5~cm$^2$ SSD planes (25~$\mu$m pitch strips in the central 1~cm,
50~$\mu$m beyond) were located downstream of the target region. These
were followed by eight 5$\times $5~cm$^2$ SSD planes of 50~$\mu$m
pitch.  The analysis dipole magnet imparted a $0.45~{\rm GeV}/c$ $p_T$
impulse in the horizontal plane to charged particles.  Downstream
track segments were measured by means of four stations of four views
($XYUV$) of 2.54~mm pitch PWCs and two stations of eight (4$X$4$Y$)
layers of STDCs with tube diameters 1.03~cm (upstream station) and
1.59~cm (downstream station)~\cite{E706-STDC}.

\begin{figure}
\epsfxsize=4in
\vskip1.3cm
\epsfbox[0 72 612 720]{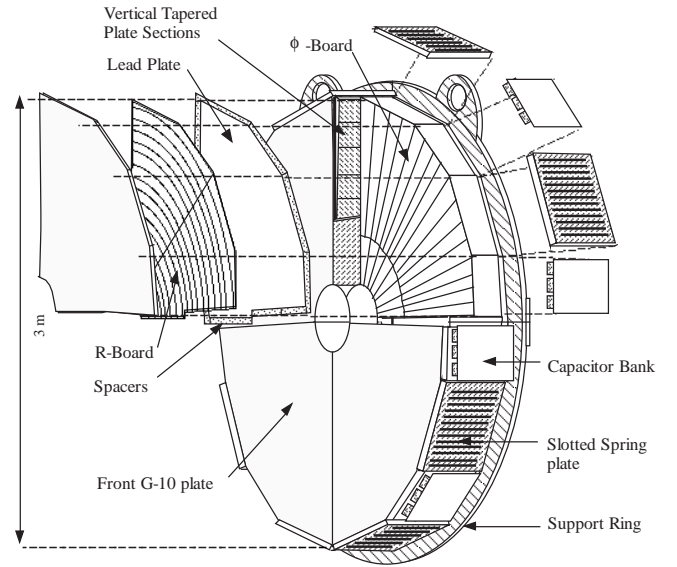}
\vskip-4.0cm
\caption{A drawing of the
liquid argon electromagnetic calorimeter with some components
pulled away in one quadrant to reveal a view of the internal details.}
\label{fig:emlac}
\end{figure}

Photons were detected in a large, lead and liquid-argon sampling
electromagnetic calorimeter (EMLAC), located 9~m downstream of the
target~\cite{E706-calibration}.  The EMLAC had a cylindrical geometry
with an inner radius of 20~cm and an outer radius of 160~cm.  The
calorimeter had 33~longitudinal cells read out in two sections: an 11
cell front section (8.5~radiation lengths) and a 22 cell back section
(18~radiation lengths).  Each longitudinal cell consisted of a 2~mm
thick lead cathode (the first cathode was constructed of aluminum), a
double-sided copper-clad G-10 radial ($R$) anode board, a second 2~mm
thick lead cathode, and a double-sided copper-clad G-10 azimuthal
($\Phi$) anode board.  The 2.5~mm gaps between these layers were
filled with liquid argon.  The physical layout is illustrated in
Fig.~\ref{fig:emlac}.

%%%%%%%%%%%%%%
%%%%%%%%%%%%%%

The EMLAC readout was subdivided azimuthally into octants, each
consisting of interleaved, finely segmented, radial and azimuthal
views.  This segmentation was realized by cutting the copper-cladding
on the anode boards to form either radial or azimuthal strips.
Signals from corresponding strips from all $R$ (or $\Phi$) anode
boards in the front (or back) section of a given octant were jumpered
together.  The copper-cladding on the radial anode boards was cut into
concentric strips centered on the nominal beam axis.  The width of the
strips on the first $R$ board was 5.5~mm.  The width of the strips on
the following $R$ boards increased slightly so that the radial
geometry was projective relative to the target region.  The azimuthal
strips were split at a radius of 40~cm into inner and outer segments;
each inner strip subtended an azimuthal angle of $\pi/192$~radians,
while outer strips covered $\pi/384$~radians.

%The overall apparatus also included two other calorimeters: a hadronic
%calorimeter located downstream of the EMLAC within the same cryostat,
%and a steel and scintillator calorimeter positioned further downstream
%to increase coverage in the very forward region.
The spectrometer was located at the end of the Meson West beamline.
The design of the beamline, primary target, and primary beam dump were
intended to minimize the rate of beam-halo muons incident upon the
spectrometer.  The beamline was capable of transporting either a
primary (800~GeV/$c$) proton beam or unseparated secondary particle
beams of either polarity to the experimental hall.  The beamline
\v{C}erenkov detector was used to identify the secondary beam
particles~\cite{striley}.  This 43.4~m long helium-filled counter was
located 100~m upstream of the experimental target.  The positive
secondary beam with mean momentum of 530~GeV/$c$ was 97\% protons.
The negative secondary beam with mean momentum of 515~GeV/$c$ was 99\%
pions.

At the end of the beamline was a 4.7~m long stack of steel surrounding
the beam pipe and shadowing the EMLAC to absorb off-axis hadrons.  A
water tank was placed at the downstream end of this hadron shield to
absorb low-energy neutrons. Surrounding the hadron shield and neutron
absorber were walls of scintillation counters (VW) to identify
penetrating muons.  There was one wall at the upstream end and two
walls at the downstream end of the hadron absorber during the 1990
run. An additional wall was added to the upstream end of the hadron
absorber prior to the 1991--1992 run.

\subsection{Trigger}

The E706 trigger selected interactions yielding high-$p_T$ showers in
the EMLAC. The selection process involved several stages: beam and
interaction definitions, a pretrigger, and high-$p_T$ trigger
requirements~\cite{E706-trigger,E706-charm,E706-pos-pieta}.  A
scintillator hodoscope, located 2~m upstream of the target region, was
used to detect beam particles, and reject interactions with more than
one spacially isolated incident particle.  Additional scintillator with 
a 1~cm diameter
central hole was located just downstream of the beam hodoscope, and
served to reject interactions initiated by particles in the beam
halo~\cite{BH}.  Two pairs of scintillator counters, mounted on the
dipole analysis magnet, were used to identify interactions in the
target.  To minimize potential confusion in the EMLAC due to
out-of-time interactions, a filter was employed 
to reject interactions that occurred within 60~ns of one
another.

For those interactions that satisfied the beam and interaction
requirements, the $p_T$ deposited in various regions of the EMLAC was
evaluated by weighting the energy signals from the EMLAC $R$-channel
amplifier fast outputs by a factor proportional to $\sin{\theta_i}$,
where $\theta_i$ was the polar angle between the $i^{th}$ strip and
the nominal beam axis.  The \PRETH\ requirement for a given octant was
satisfied when the $p_T$ detected in either the inner 128 $R$ channels
or the outer $R$ channels of that octant was greater than a threshold
value.  A pretrigger signal was issued only when there was no evidence
in that octant of substantial noise, significant $p_T$ attributable to
an earlier interaction, or incident beam-halo muon detected by the VW.

Localized trigger groups were formed for each octant by clustering the
$R$-channel fast-outputs into 32~groups of 8~channels.  Each adjacent
pair of 8~channel groups formed a group-of-16 strips.  If the $p_T$
detected in any of these groups-of-16 was above a specified high (or
low) threshold, then a \LOCALH\ (or \LOCALL) signal was generated for
that octant.  A \SLH\ (or \SLL) trigger was generated if a \LOCALH\
(or \LOCALL) signal was generated in coincidence with the \PRETH\ in
the same octant.

Trigger decisions were also made based upon global energy depositions
within an octant.  A \GLOBALL\ signal was generated if the total $p_T$
in an octant exceeded a threshold value.  The \GLL\ trigger required a
coincidence of the \PRETH\ signal with \GLOBALL\ and \LOCALL\ signals
from the same octant. The \LOCALL\ requirement was included to
suppress spurious global triggers due to coherent noise in the EMLAC.

The \SLL\ and \GLL\ triggers were prescaled to keep them from
dominating the trigger rate. Prescaled samples of beam, interaction, and 
pretrigger events were also recorded.

\section{Analysis Methods}

Data samples contributing to this analysis represent an integrated
luminosity of~1.6 (6.8)~pb$^{-1}$ and 1.6 (6.5)~pb$^{-1}$ for
530 and 800~GeV/$c$ $p$Cu ($p$Be) interactions, respectively, as well as
0.3 (1.4)~pb$^{-1}$ for 515~GeV/$c$ $\pi^-$Cu ($\pi^-$Be) interactions. 
These samples
were accumulated during the 1991--1992 run.  Results reported in this
paper also use 0.9 (6.1)~pb$^{-1}$ of $\pi^-$Cu ($\pi^-$Be) data recorded 
during the 1990 run.
The following subsections describe the analysis procedures and methods
used to correct the data for losses from inefficiencies and
selection biases.  Additional details can be found in our previous
papers~\cite{E706-pos-pieta,E706-neg-pieta,E706-charm,E706-directphoton,
E706-calibration}.

\subsection{Charged-particle reconstruction}

The two major aspects of the analysis procedure involved
charged-particle and calorimeter-shower reconstruction (discussed in
Sec.~\ref{sec:calorimeter}).  The charged-track reconstruction algorithm produced
track segments upstream of the magnet using information from the SSDs,
and downstream of the magnet using information from the PWCs and
STDCs. These track segments were projected to the center of the magnet
and linked to form final tracks whose calculated charges and momenta
were used for the physics analysis.  The charged track reconstruction
is described in more detail elsewhere~\cite{E706-charm,blusk}.
The primary vertex reconstruction is described in 
Sec.~\ref{sec:vertex}.

\subsection{Calorimeter shower reconstruction}
\label{sec:calorimeter}

The readout of each EMLAC quadrant was divided into four regions: left
and right $R$, and inner and outer~$\Phi$.  Strip energies from
clusters in each region were fit to the shape of an electromagnetic
shower determined from detailed Monte Carlo simulations and
isolated-shower data.  These fits were used to evaluate the positions
and energies of the peaks in each region.  Shower positions and
energies were obtained by correlating peaks of approximately the same
energy in the $R$ and $\Phi$ regions within the same half octant.
More complex algorithms were used to handle configurations with
showers spanning multiple regions.  The EMLAC readout was also
subdivided longitudinally into front and back sections.  This
segmentation provided discrimination between showers generated by
electromagnetically or hadronically interacting particles. Photons
were defined as showers with
at least 20\% of the shower energy deposited in
the front part of EMLAC, to reduce the backgrounds due to showers from
hadronic interactions.  Losses of photons due to this requirement were
$\approx2$\%.  A detailed event simulation was employed to correct for
this and other effects including reconstruction smearing and losses.
An expanded discussion of the EMLAC reconstruction procedures and
performance can be found elsewhere~\cite{E706-calibration}.

\subsection{Meson signals}

For this study, $\pi^0$ and $\eta$ mesons were reconstructed via their
$\gamma\gamma$ decay modes.  Only those $\gamma\gamma$ combinations
with energy asymmetry $A_{\gamma\gamma} = |E
_{\gamma_1}-E_{\gamma_2}|/ (E_{\gamma_1}+ E_{\gamma_2})<0.75$ were
considered to reduce uncertainties due to low energy photons. The
meson signals have been corrected for losses due to the energy
asymmetry cut and the branching fractions for the $\gamma\gamma$ decay
modes~\cite{pdg}.

Photons were required to be reconstructed within the fiducial region
of the EMLAC to exclude areas with reduced sensitivity.  In addition,
$\gamma\gamma$ combinations were restricted to the same octant to
simplify the trigger analysis.  A simple ray-tracing Monte Carlo
program was employed to determine the correction for these fiducial
requirements.

The correction for losses due to the conversion of photons into
$e^+e^-$ pairs was evaluated by projecting each reconstructed photon
from the event vertex to the reconstructed position in the EMLAC.  The
radiation length of material traversed, up to the analysis magnet, was
evaluated based upon detailed detector descriptions.  The photon
conversion probability was evaluated and used to account for
conversion losses. The average correction for conversion losses was
$1.09$ per photon for the Be target in the 1990 run ($1.08$ in
1991--1992 run) and $1.19$ per photon for the Cu target ($1.16$ in
1991--1992 run).

\subsection{Detector simulation}

The Meson West spectrometer was modeled with a detailed {\sc
geant}~\cite{geant} simulation (DGS).  A preprocessor was used to
convert {\sc geant} information into the simulated hits and strip energies
associated with the various detectors.  The preprocessor simulated
hardware effects, such as channel noise and gain variations.  Monte
Carlo events were then processed through the same reconstruction
software used for the analysis of the data.  This technique accounted
for inefficiencies and biases in the reconstruction algorithms.
Reconstuction inefficiencies were relatively small over most of the
kinematic range. More information on the detailed simulation of the Meson
West spectrometer can be found
elsewhere~\cite{E706-directphoton,E706-pos-pieta,apana}.

%%%%%%%%%%%%%%
\begin{figure}
\epsfxsize=\figsize
\epsfbox[\bbcoord]{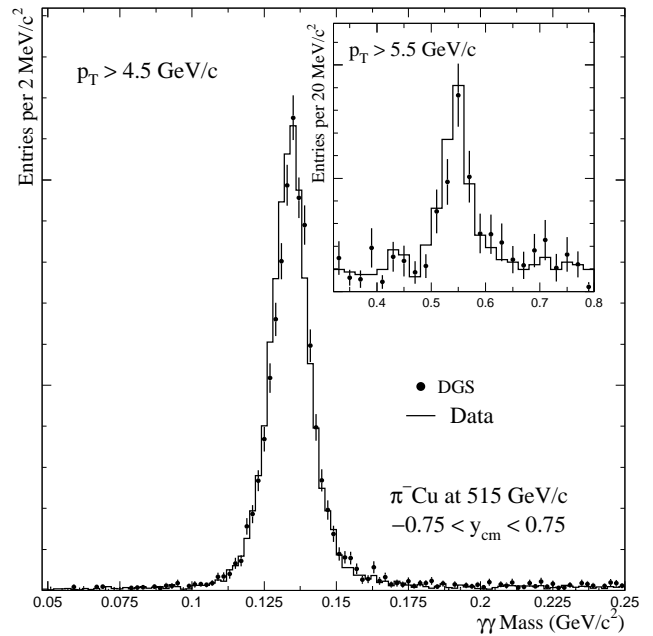}
\caption{Comparison between data~(histogram) and the detailed Monte Carlo 
simulation ($\bullet$) for 515~GeV/$c$ $\pi^-$ beam on Cu for
$\gamma\gamma$ combinations in the $\pi^0$ and $\eta$ mass regions.}
\label{fig:mc-pi0-eta_cu}
\end{figure}
%%%%%%%%%%%%

\begin{figure}
\epsfxsize=\figsize
\epsfbox[\bbcoord]{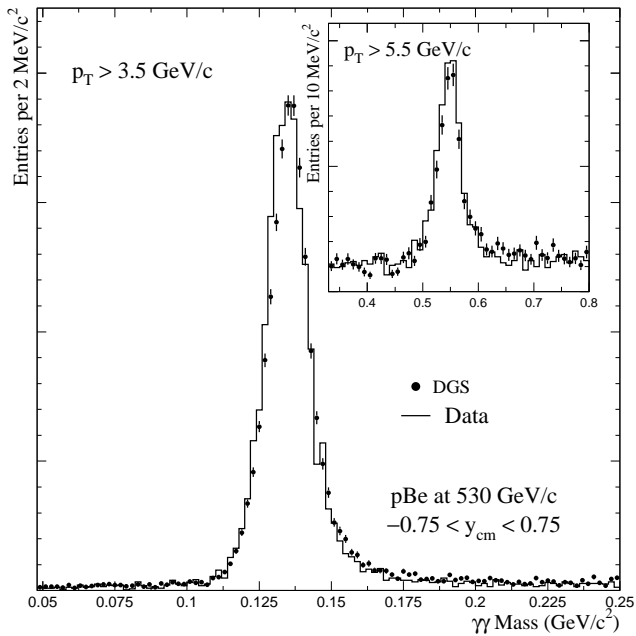}
\caption{Comparison between data~(histogram) and the detailed Monte Carlo 
simulation ($\bullet$) for 530~GeV/$c$ proton beam on Be for
$\gamma\gamma$ combinations in the $\pi^0$ and $\eta$ mass regions.}
\label{fig:mc-pi0-eta_be}
\end{figure}

\begin{figure}
\epsfxsize=\figsize
\epsfbox[\bbcoord]{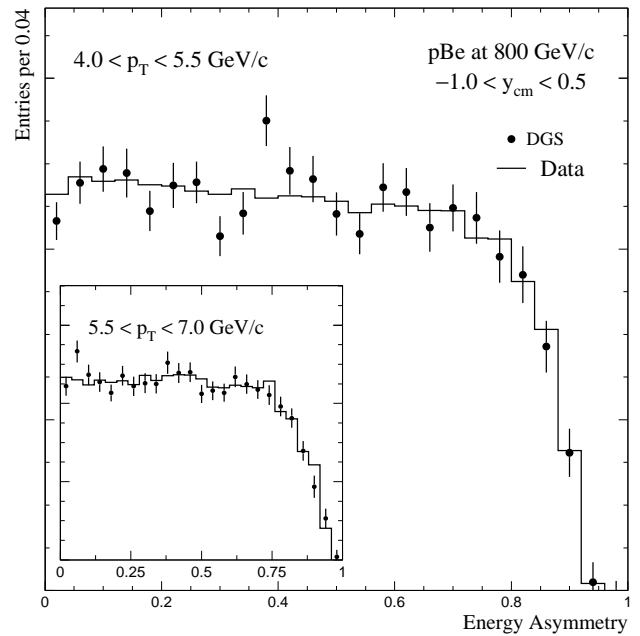}
\caption{Comparison of $A_{\gamma\gamma}$, the energy asymmetry 
distribution for photons from $\pi^0$ decays, in data
(histogram) and the detailed Monte Carlo simulation ($\bullet$) for the
800~GeV/$c$ proton beam sample.  Shown are the comparisons for two $p_T$
intervals, $4.0<p_T< 5.5$~GeV/$c$ and $5.5<p_T<7.0$~GeV/$c$.}
\label{fig:mc-A}
\end{figure}

As inputs to the {\sc geant} simulation, we employed single particle
distributions, reconstructed data events, and {\sc
herwig}-generated~\cite{herwig56} events.  The {\sc herwig}-generated
$\pi^0$, $\eta$, and direct-photon spectra were weighted in $p_T$ and
rapidity to our measured results in an iterative fashion so that the
final corrections were based on the data distributions rather than on
the behavior of the physics generator.  Figure~\ref{fig:mc-pi0-eta_cu}
shows the $\gamma\gamma$~mass spectra in the $\pi^0$ and $\eta$
mass regions in comparison to the DGS results for the $\pi^-$Cu data
at 515 GeV/$c$, and Fig.~\ref{fig:mc-pi0-eta_be} shows an analogous
plot for our higher statistics $p$Be data at 530 GeV/$c$.
In addition to providing evidence that
the DGS simulated the EMLAC resolution well, the agreement between the
levels of combinatorial background indicates that the DGS also
provided a reasonable simulation of the underlying event structure.
Since the DGS was tuned using our higher statistics Be data, 
%spaz
Figs.~\ref{fig:mc-A}--\ref{fig:pmc} also show the level of agreement
achieved for this target.
Figure~\ref{fig:mc-A} shows a comparison between the DGS and the data
for the sideband-subtracted energy asymmetry distribution for photons
from $\pi^0$ decays.  This figure illustrates that the
simulation accurately describes the losses of low-energy photons.

%%%%%%%%%%%%%%
%%%%%%%%%%%%

%%%%%%%%%%%%%%
%%%%%%%%%%%%

A second Monte Carlo simulation of the detector (PMC) was used to
cross check the detailed simulation and for studies that required
large statistics.  This simulation employed parameterizations of
physics cross sections and detector
responses~\cite{E706-directphoton,begel,apana}.  The inclusive $\pi^0$ and
direct-photon cross sections were parameterized as two dimensional
surfaces in $p_T$ and rapidity~\cite{apana}.  The $\eta$, $\omega$,
and $\eta^\prime$ cross sections were parameterized using the measured
$\eta/\pi^0$~\cite{E706-pos-pieta,E706-neg-pieta},
$\omega/\pi^0$~\cite{E706-omega}, and
$\eta^\prime/\eta$~\cite{etaprimeratio,geist} ratios.  Generated
mesons were decayed into final state particles; photons were smeared
for energy and position resolution~\cite{E706-calibration}.  A vertex
was generated in the simulated target for every event.  Photons were
allowed to convert into $e^+e^-$ pairs; the energy of the resulting
electrons was reduced using the {\sc geant} function for
bremsstrahlung radiation.  Electron four-vectors were smeared for
multiple scattering in the target and the resolution of the tracking
system and adjusted for the magnet impulse.
Figure~{\ref{fig:pmc} displays a comparison between the PMC and the
data in the $\pi^0$ and $\eta$ mass regions and for the $\pi^0$ energy
asymmetry.  The PMC provides an adequate characterization of the data.
%%%%%%%%%%%%%%
\begin{figure}
\epsfxsize=\figsize
\epsfbox[\bbcoord]{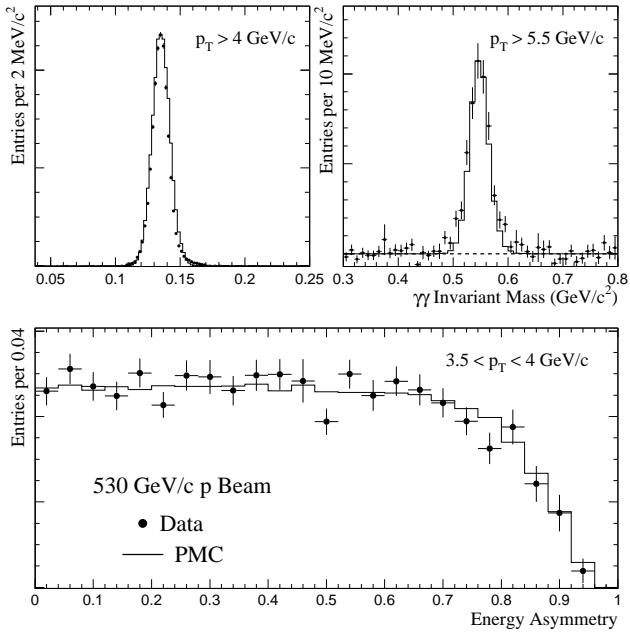}
\caption{Comparison between data ($\bullet$) and the parameterized Monte
Carlo (histogram) from the 530~GeV/$c$ proton beam sample: (top)
$\gamma\gamma$ combinations in the $\pi^0$ and $\eta$ mass regions;
(bottom) asymmetry in energy for photons from $\pi^0$ mesons.  The 
combinatorial background in
the data has been removed from this comparison through a simple
subtraction. The distributions have been normalized to the same area.}
\label{fig:pmc}
\end{figure}
%%%%%%%%%%%%

\subsection{Vertex reconstruction}
\label{sec:vertex}

The location of the interaction vertex was reconstructed using
charged-particle tracks.  Vertices were identified by means of an
impact-parameter minimization technique~\cite{blusk}.  A $\chi^2$ was
defined for a given vertex position using the impact parameters of the
reconstructed tracks and their projection uncertainties.  The vertex
position was found by minimizing this $\chi^2$.  Vertices were found
in~$X$ and~$Y$ independently and correlated based on the
difference in their positions along the nominal beam direction ($Z$
axis).  The $Z$ position of the matched vertex was the weighted
average of the $Z$ positions found in~$X$ and~$Y$.  
%{\bf In cases where several matched
%vertices were found, the vertex located furthest downstream was
%assumed to be the primary vertex??? CHECK PLREC.} 
The reconstructed vertex positions are presented in
Fig.~\ref{fig:cmp_vz} as functions of $Z$ for the two target
configurations.  The beryllium, copper, and hydrogen targets are
clearly visible, as are the SSDs and related support structures.  The
average resolution for the $Z$ location of the interaction vertex was
$\approx300\,\mu$m~\cite{E706-charm,blusk}.

%%%%%%%%%%%%%%
\begin{figure}
\epsfxsize=\figsize
\epsfbox[\bbcoord]{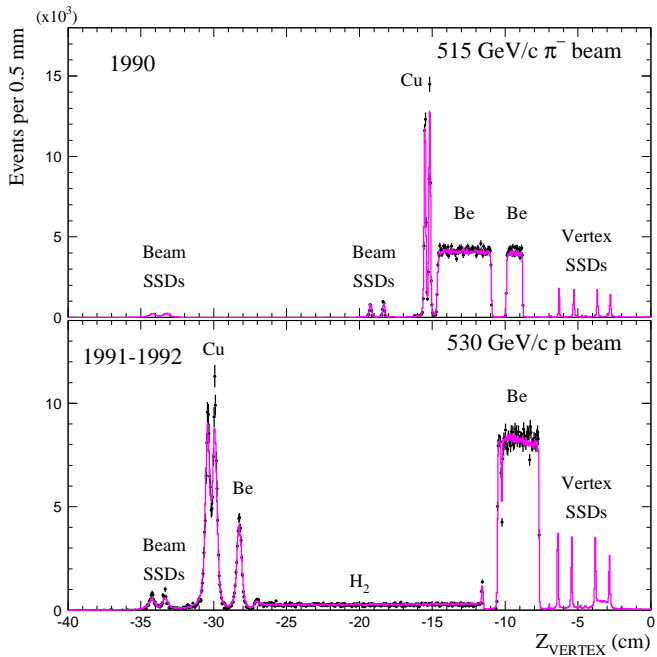}
\caption{$Z$ positions of primary vertices in the detailed 
Monte Carlo simulation ($\bullet$) and the data (histogram) for events
containing a $\pi^0$ candidate with $p_T>4$ GeV/$c$.}
\label{fig:cmp_vz}
\end{figure}
%%%%%%%%%%%%
%%%%%%%%%%%%%%
\begin{figure}
\epsfxsize=\figsize
\epsfbox[\bbcoord]{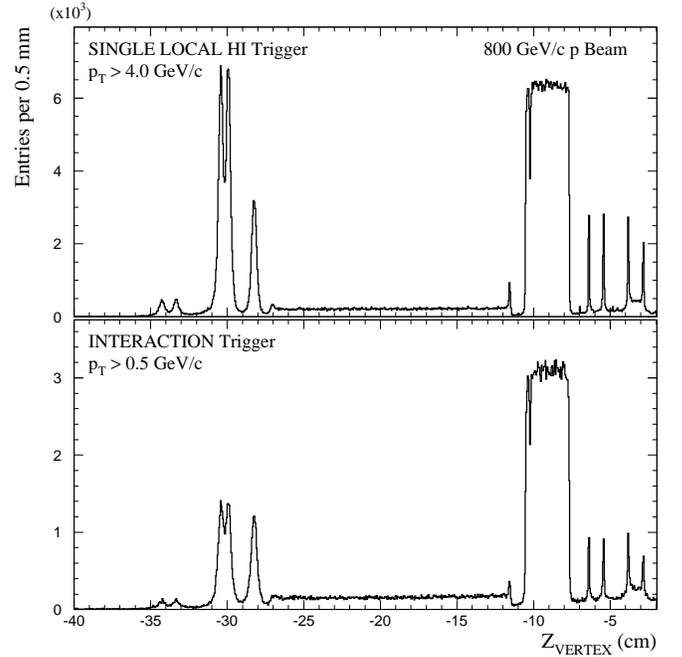}
\caption{$Z$ positions of primary vertices for events containing
$\pi^0$ candidates with $p_T>4.0$~GeV/$c$ acquired using the
{\sc single local hi} trigger (top) and for events containing
$\pi^0$ candidates with $p_T>0.5$~GeV/$c$ 
acquired using the {\sc interaction} trigger (bottom).}
\label{fig:vz_trigger}
\end{figure}
%%%%%%%%%%%%
%%%%%%%%%%%%%%
\begin{figure}
\epsfxsize=\figsize
\epsfbox[\bbcoord]{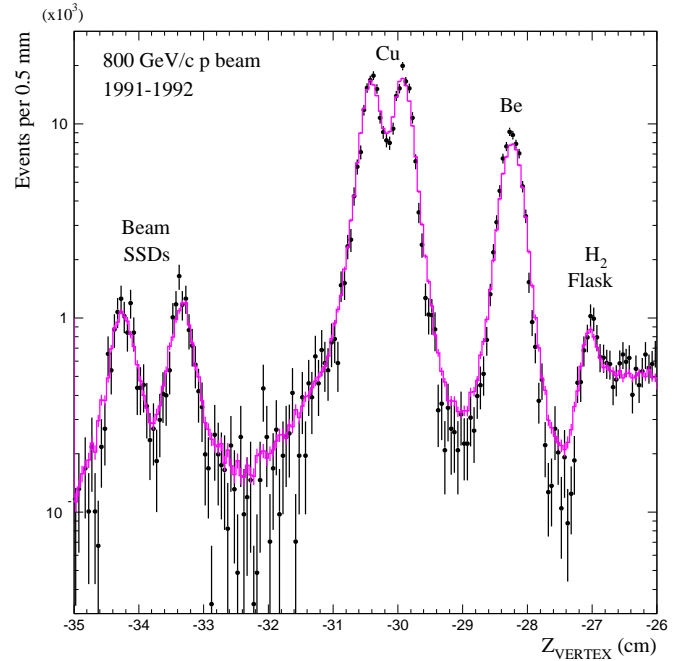}
\caption{$Z$ positions of primary vertices in the vicinity
of the Cu targets in the detailed
Monte Carlo simulation ($\bullet$) and the data (histogram), 
for events containing $\pi^0$ candidates with $p_T>4.0$~GeV/$c$ in the
800~GeV/$c$ sample. Note the use of logarithmic scale.}
\label{fig:vz_cu_be}
\end{figure}
%%%%%%%%%%%%
The relative heights of the Cu and Be targets shown in
Fig.~\ref{fig:cmp_vz} varies as a function of $p_T$.  This is clearly
evident in Fig.~\ref{fig:vz_trigger}, which compares two $\pi^0$
samples: one acquired using the highly prescaled {\sc interaction}
trigger, and the other using the {\sc single local hi} trigger.  The
{\sc interaction} triggered events are typically minimum-bias in
character with low-$p_T$ $\pi^0$'s.  The number of primary vertices 
scales as $\approx A^{2/3}$, where $A$ is the atomic weight of the target.  
The {\sc single local hi} triggers are typically caused by hard-scatters
that produce high-$p_T$ $\pi^0$'s.  The number of primary vertices 
in these events scale as $\approx A^1$.

The DGS was used for detailed studies of the vertex reconstruction.
The transverse positions of vertices were chosen according to beam
profiles observed in the data.  Longitudinal positions were determined
using Monte Carlo methods based upon the interaction lengths of the
materials in the target region (Table~\ref{tab:target}).  DGS events
were weighted to reproduce the relative number of vertices in the
data.  Results from the DGS compare favorably with the data in
Fig.~\ref{fig:cmp_vz}.  This good agreement was particularly important
for separating events with primary vertices in copper from those
in the upstream piece of beryllium.  Figure~\ref{fig:vz_cu_be}
displays the longitudinal vertex distribution, focussing on this region
in the 1991--1992 target configuration.  The shape of the tails in the
data are well described by the DGS.

The vertex reconstruction efficiency was evaluated using the
DGS~\cite{apana}.  Separate reconstruction efficiencies were evaluated
for the Be, Cu, and H${}_2$ targets.  The reconstruction probability
was defined as the number of vertices reconstructed in each target's
fiducial volume divided by the number of vertices generated in the
fiducial volume.  The reconstruction efficiency was the inverse of
this probability.  Defined in this manner, the reconstruction
efficiency also corrected for the longitudinal resolution smearing of
reconstructed vertices.

Additional beam particles occasionally interacted in the target
material within the data-capture timing window of the tracking system.
The extra tracks sometimes caused the vertex associated with the
high-$p_T$ interaction to be misidentified.  The bias introduced by
these rare events favored configurations where the low-$p_T$
interaction took place within the downstream piece of Be. This
primarily affected interactions in the Cu and upstream Be targets in
the long 1991--1992 target configuration because of the relatively poor
vertex resolution in those targets compared to the downstream Be.
This bias was investigated by comparing the $\pi^0$ cross sections
measured in $\pi^-$Be interactions in the 1990 and the 1991--1992 runs,
and by comparing $\pi^0$ yields from the upstream and downstream Be
pieces in the 1991--1992 target configuration.  The number of Cu
vertices were corrected for misidentifications arising from this
source.  The resulting correction was $1.04$ for the 1991--1992 $\pi^-$Cu
sample at 515~GeV/c, $1.06$ for the 530~GeV/$c$ $p$Cu sample, and 
$1.12$ for the 800~GeV/$c$ $p$Cu sample~\cite{hydrogen-correction}.

Each event in this analysis was required to have a reconstructed
vertex in the target region.  Longitudinal and transverse requirements
were placed on vertices to define the data samples.  The
longitudinal cuts selected the target in which the incident beam
interacted, while the transverse cuts ensured that the interaction
occurred within the target material.

\subsection{Direct photons}

The largest contribution to the direct-photon background comes from
electromagnetic decays of neutral hadrons, particularly $\pi^0$'s and
$\eta$'s.  For the purposes of the measurements reported here, a
photon was a direct-photon candidate if it did not combine with
another photon in the same octant to form a~$\pi^0$ with 
$A_{\gamma\gamma} \leq 0.9$ or an $\eta$~with $A_{\gamma\gamma} \leq 0.8$.  

To suppress electrons, reconstructed showers were excluded from the
sample when charged-particle tracks pointed to within 1~cm of shower
center.  The correction for this criterion in the direct-photon
analysis is $\approx1.01$ based upon studies of the impact of this
requirement on reconstructed $\pi^0$'s.

The residual background from $\pi^0$'s and $\eta$'s, as well as from
other sources of background, was calculated using DGS samples that
contained no generated direct photons ($\gamma_b$). A smooth fit of the
$\gamma_b/\pi^0$ in $p_T$ and rapidity was used to extract the
direct-photon cross sections.  The systematic uncertainty in this
background subtraction was estimated by varying the direct-photon
definition as a function of the cut on $A_{\gamma\gamma}$.
The direct-photon background was
also investigated using the PMC.  Figure~\ref{fig:gpi_pmc} compares
the direct-photon backgrounds estimated using the two Monte Carlo
simulations.  The close agreement provides additional confidence in our
understanding of the background in our direct photon samples. Additional
details are provided in Ref.~\cite{E706-directphoton}.
%%%%%%%%%%%%%%
\begin{figure}
\epsfxsize=\figsize
\epsfbox[\bbcoord]{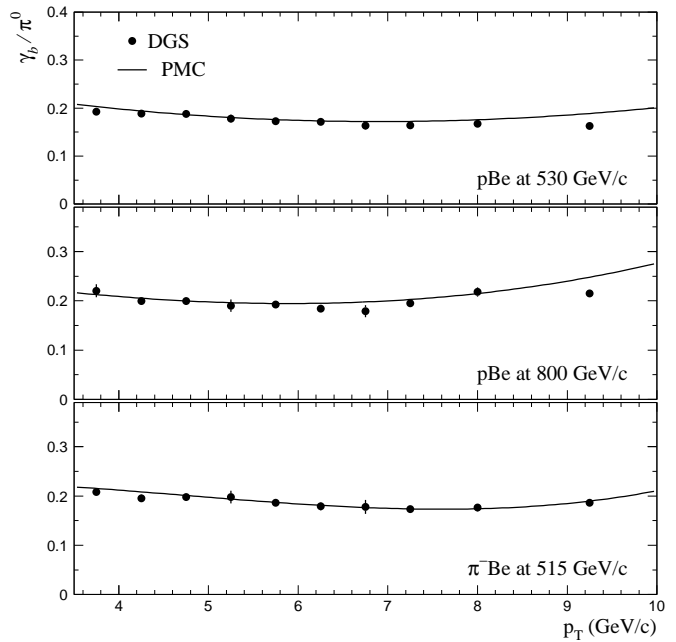}
\caption{A comparison of the direct-photon background as a function of
$p_T$ as predicted by the parameterized Monte Carlo simulation (line)
and the detailed Monte Carlo simulation ($\bullet$).  The background
was normalized to the $\pi^0$ cross section.}
\label{fig:gpi_pmc}
\end{figure}
%%%%%%%%%%%%

Fits to $\gamma_b/\pi^0$ were only made for the beryllium target
due to relatively poor DGS statistics in the other targets. However,
$\gamma_b/\pi^0$ is expected to be slightly different for each target
due to the different amounts of target material the photons must
traverse. Therefore, a correction to $\gamma_b/\pi^0$ was calculated
using the PMC. The differences in $\gamma_b/\pi^0$ are shown for the
800~GeV/$c$ proton beam data in Fig.~\ref{fig:tgt_fix}.  The other
data samples have similar behavior.  The target differences were fit as
functions of $p_T$ and rapidity for each incident beam and target
configuration and applied as
additive corrections to the nominal $\gamma_b/\pi^0$ fit.
%%%%%%%%%%%%%%
\begin{figure}
\epsfxsize=\figsize
\epsfbox[\bbcoordb]{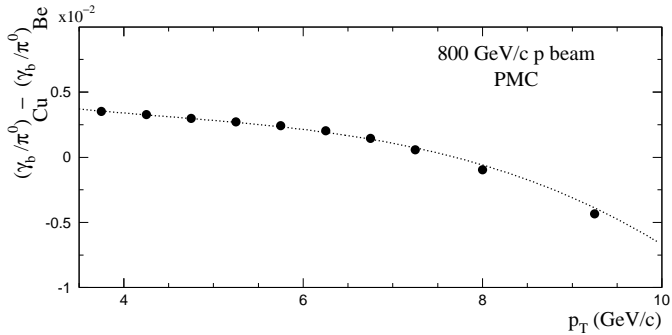}
\caption{Difference between the
copper and beryllium background $\gamma/\pi^0$ ratios versus $p_T$ for
the 800~GeV/$c$ proton beam. The dotted
line represents a fit to the difference integrated over rapidity.}
\label{fig:tgt_fix}
\end{figure}
%%%%%%%%%%%%

The impact of the background can be determined by normalizing the
direct-photon candidate spectrum from the data and the simulation to
the measured $\pi^0$ cross
section~\cite{E706-pos-pieta,E706-neg-pieta}.  This $\gamma/\pi^0$
ratio is displayed in Fig.~\ref{fig:gpi} for all three incident beams
for the copper target.  The signal-to-background in all cases is large
at high $p_T$.

%%%%%%%%%%%%%%
\begin{figure}
\epsfxsize=\figsize
\epsfbox[\bbcoord]{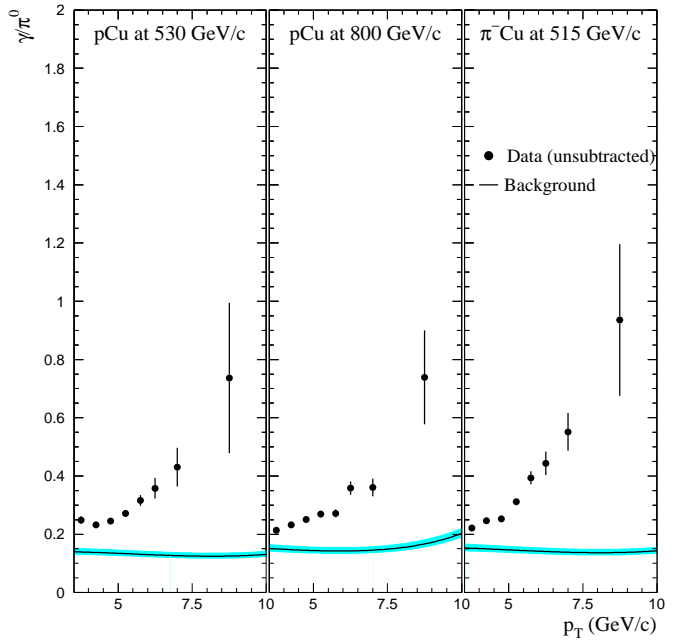}
\caption{$\gamma/\pi^0$ ratio for each of the three beam samples as a
function of $p_T$.  Direct-photon candidates in the data are indicated
by the points. The error bars represent statistical contributions to the
uncertainties. The line indicates the expected background and its
width is representative of the associated systematic uncertainty.}
\label{fig:gpi}
\end{figure}
%%%%%%%%%%%%

\subsection{Summary of systematic uncertainties}

The systematic uncertainties for the production of $\pi^0$'s,
$\eta$'s, and direct photons measured in $\pi^-$Cu and $p$Cu
interactions are similar to those detailed in
Refs.~\cite{E706-pos-pieta,E706-neg-pieta,E706-directphoton}.  The
principal contributions to the systematic uncertainty arose from the
following sources: normalization, calibration of photon energy
response and detector-resolution unsmearing, background subtraction,
reconstruction efficiency, incident beam contamination (for the
515~GeV/$c$ and 530~GeV/$c$ secondary beams), beam halo muon
rejection, geometric acceptance, photon conversions, trigger response,
and vertex finding.  The additional vertex-finding uncertainty
associated with the confusion induced by multiple beam particles
interacting in the target was $\approx2\%$.  The total systematic
uncertainties, combined in quadrature, are quoted with the cross
sections in the appropriate tables.  Note that some of these
contributions to the systematic uncertainty (e.g. normalization) are
strongly correlated between bins.

Most of the experimental systematic uncertainties cancel in the ratio
of cross sections measured on different targets.  The residual
uncertainties are due to target-related systematics associated mainly
with vertex identification.  The total systematic uncertainty in the
ratio of cross sections measured on Cu to those on Be is $\pm3\%$ for
the 800~GeV/$c$ $p$ beam sample, and $\pm2\%$ for the 530~GeV/$c$ $p$
and 515~GeV/$c$ $\pi^-$ beam samples.  The systematic uncertainty
associated with the ratios of Be to H and Cu to H are $\pm4\%$ for all
samples after correcting for the effects of vertex misidentification
as discussed in Sec.~\ref{sec:vertex}.

%{\bf Shifts in Be and H data from previous papers must be indicated.}

\section{Results and Discussion}

\subsection{Cross sections}

The invariant differential cross sections per nucleon for
direct-photon, $\pi^0$, and $\eta$ production from 530 and 800~GeV/$c$ 
$p$ beams and 515~GeV/$c$ $\pi^-$ beam incident on copper are
presented as functions of $p_T$ in Figs.~\ref{fig:xs_530_cu}
through~\ref{fig:xs_515_cu}.  Results from 530~GeV/$c$ $p$ 
and 515~GeV/$c$ $\pi^-$ beams are averaged over the rapidity range
$-0.75\le\ycm\le 0.75$; results from the 800~GeV/$c$ $p$ beam are
averaged over $-1.0\le\ycm\le 0.5$.  Data points are plotted at
abscissa values that correspond to the average value of the cross
section in each $p_T$ bin, assuming local exponential $p_T$
dependence~\cite{laff}.  The inclusive cross sections are tabulated in
the Appendix.
%Tables~\ref{tab:foo} and~\ref{tab:bar}.  Cross section measurements,
%binned in both $p_T$ and \ycm, are reported in
%Tables~\ref{tab:foo1}---\ref{tab:foo2}.

\begin{figure}
\epsfxsize=\figsize
\epsfbox[\bbcoord]{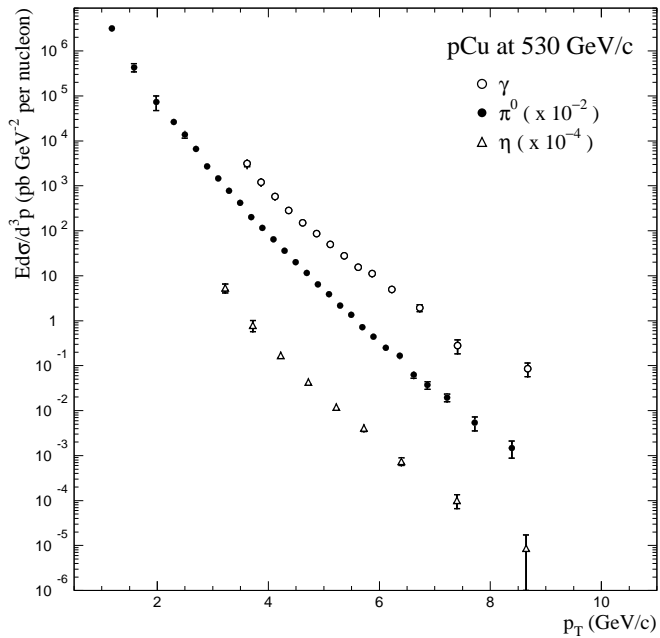}
\caption{
Invariant differential cross sections per nucleon for direct-photon,
$\pi^0$, and $\eta$ production as functions of $p_T$, averaged over
$-0.75\le\ycm\le 0.75$, from 530~GeV/$c$ proton beam incident upon a copper
target.}
\label{fig:xs_530_cu}
\end{figure}
%%%%%%%%%%%%%%
\begin{figure}
\epsfxsize=\figsize
\epsfbox[\bbcoord]{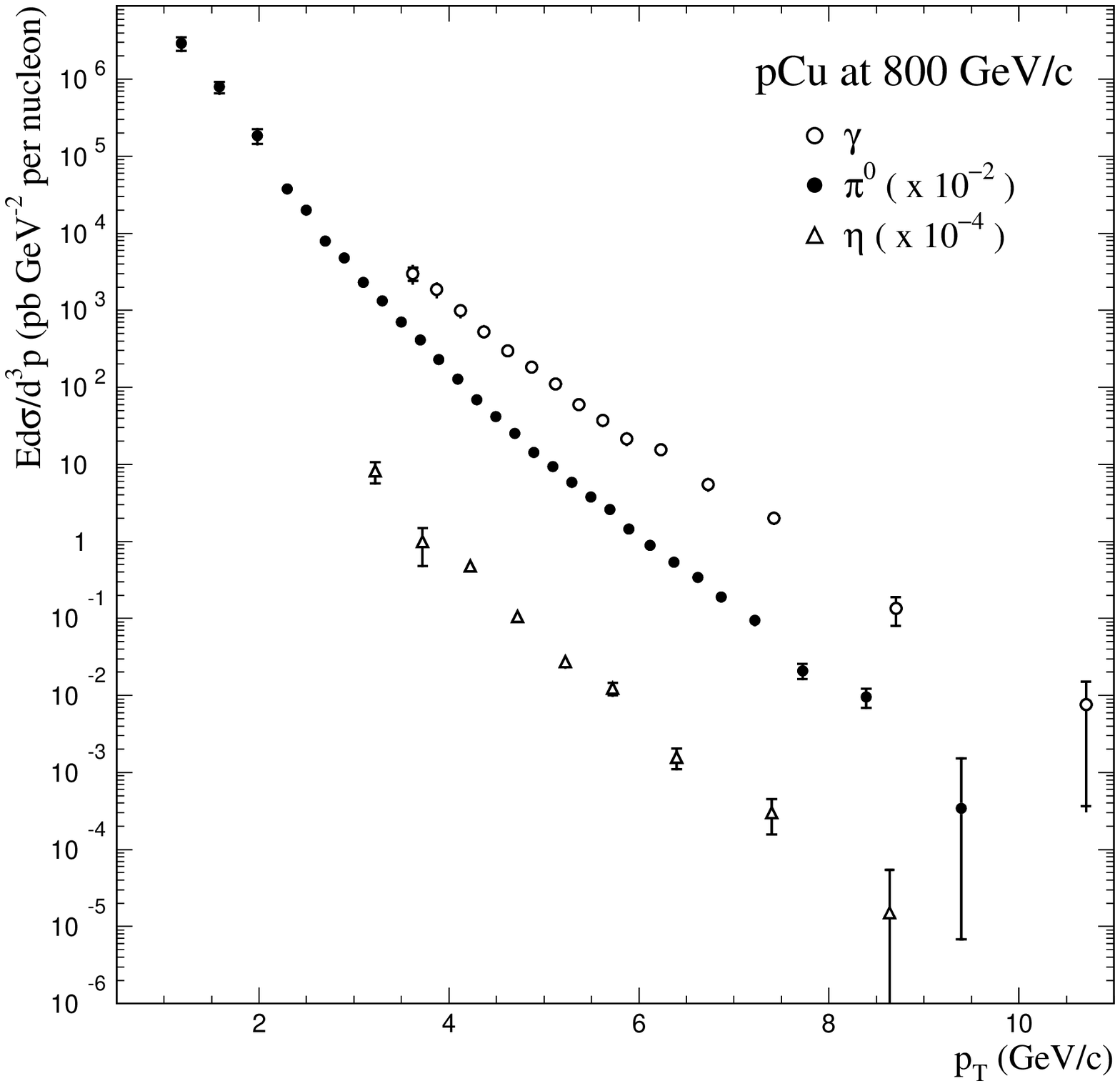}
\caption{
Invariant differential cross sections per nucleon for direct-photon,
$\pi^0$, and $\eta$ production as functions of $p_T$, averaged over
$-1.0\le\ycm\le 0.5$, from 800~GeV/$c$ proton beam incident upon a copper
target.}
\label{fig:xs_800_cu}
\end{figure}
%%%%%%%%%%%%%%
%%%%%%%%%%%%%%
\begin{figure}
\epsfxsize=\figsize
\epsfbox[\bbcoord]{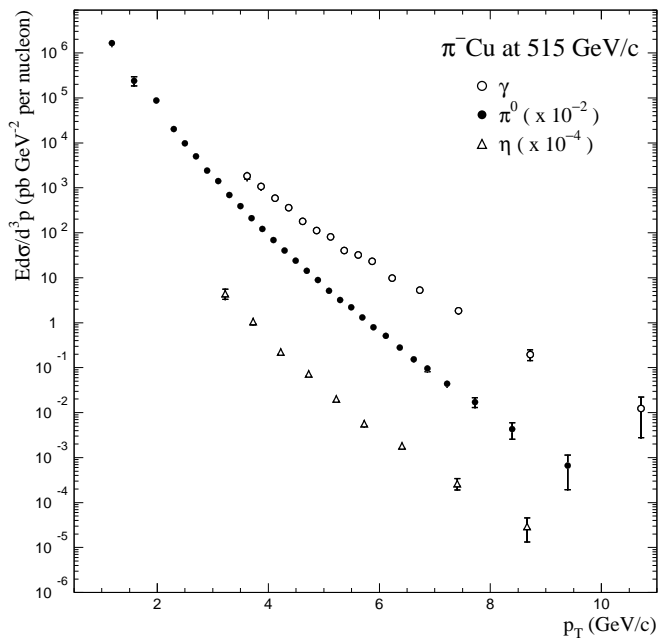}
\caption{
Invariant differential cross sections per nucleon for direct-photon,
$\pi^0$, and $\eta$ production as functions of $p_T$, averaged over
$-0.75\le\ycm\le 0.75$, for 515~GeV/$c$ $\pi^-$ beam incident upon a copper
target.}
\label{fig:xs_515_cu}
\end{figure}
%%%%%%%%%%%%%%

\subsection{Ratios}

E706 has previously reported results for direct-photon and $\pi^0$
production on beryllium and hydrogen
targets~\cite{E706-directphoton,E706-pos-pieta,E706-neg-pieta}.  Since
E706 is the only direct-photon experiment that used more than one
nuclear target, our data provide a unique measurement of nuclear
effects in direct-photon production.  Figures~\ref{fig:xs_cube_530}
to~\ref{fig:xs_cube_515} present the ratio of
inclusive cross sections per nucleon measured on Cu to those measured
on Be for $\pi^0$ mesons and direct photons.  (The
Be/H ratios were presented in previous
publications~\cite{E706-directphoton,E706-pos-pieta,E706-neg-pieta},
and are not reproduced here.)  These ratios show clear evidence of
nuclear enhancement in both $\pi^0$ and direct-photon production
and the $\pi^0$ data exhibit the decrease at high $p_T$ first noted
by Cronin~{\it et~al.}~\cite{cronin,antreasyan}. This behavior is generally 
attributed to the influence of multiple-parton scattering prior to the hard
scatter~\cite{wang-rep}.

Results from fits to constant ratios in restricted regions of $p_T$
have been overlaid on the data in Figs.~\ref{fig:xs_cube_530} 
to~\ref{fig:xs_cube_515}, and summarized in Table~\ref{tab:nuclear}. 
Fits were made for $3.5<p_T<5.5$~GeV/$c$ for the
$\pi^0$'s and over the entire $p_T$ range for the direct photons. 
The difference in the ratio
for $\pi^0$ mesons and direct photons is significant and may be
indicative of differing roles of initial and final-state effects, since
direct photons are not expected to be strongly impacted by final-state
nuclear effects.
%%%%%%%%%%%%%%
\begin{table}[b]
\caption{Results of fits of the Cu to Be ratios for $\pi^0$ and direct-photon
production.}
\begin{tabular}{l||c|c}
\multirow{2}{*}{Sample} & $\pi^0$ & direct photon \\
       &\footnotesize $3.5<p_T<5.5$ GeV/$c$ &\footnotesize $3.5<p_T<8.0$~GeV/$c$ \\
\hline\hline
530~GeV/$c$ $p$ & $1.271\pm0.016\pm0.025$ & $1.103\pm0.032\pm0.022$ \\
800~GeV/$c$ $p$ & $1.283\pm0.025\pm0.038$ & $1.043\pm0.032\pm0.031$ \\
515~GeV/$c$ $\pi^-$ & $1.237\pm0.015\pm0.025$ & $1.083\pm0.024\pm0.022$ \\
\hline\hline
\end{tabular}
\label{tab:nuclear}
\end{table}
%%%%%%%%%%%%

%%%%%%%%%%%%%%
\begin{figure}
\epsfxsize=\figsize
\epsfbox[\bbcoord]{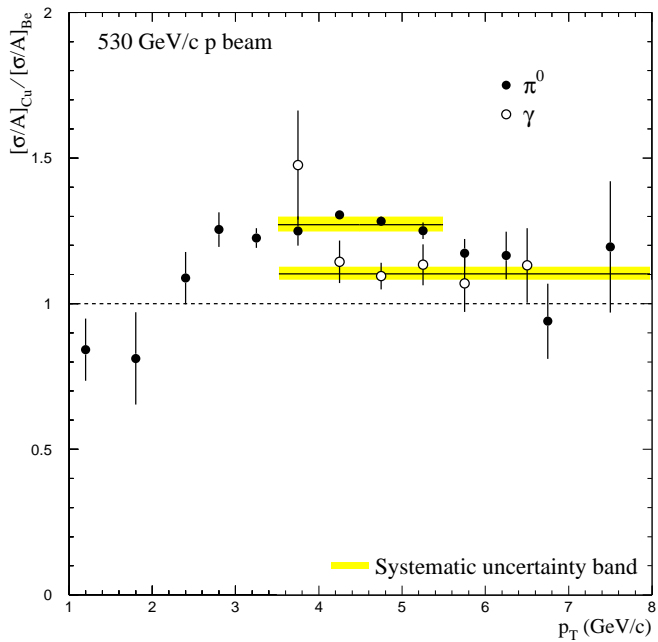}
\caption{
The ratio of inclusive $\pi^0$ and direct-photon production cross 
sections per nucleon in $p$Cu to those in $p$Be collisions at 530~GeV/$c$.
Simple straight line fits to regions with relatively flat distributions 
have been overlaid on the data.
The error bars represent only statistical contributions
to the uncertainties.  Systematic uncertainties are indicated by the
shaded region  associated with the fit.}
\label{fig:xs_cube_530}
\end{figure}
%%%%%%%%%%%%%%%%%

Expectations from a theoretical prediction for nuclear enhancement in
the 515~GeV/$c$ $\pi^-$ direct-photon sample~\cite{guo} have been
overlaid on Fig.~\ref{fig:xs_cube_515}.  This calculation predicts a
slight enhancement in direct-photon production and agrees with the data
within the uncertainties.

%%%%%%%%%%%%%%
\begin{figure}
\epsfxsize=\figsize
\epsfbox[\bbcoord]{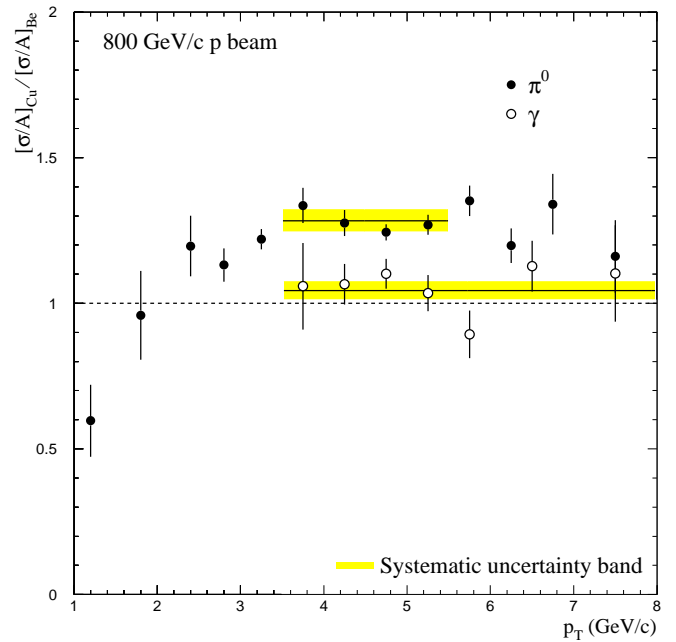}
\caption{
The ratio of inclusive $\pi^0$ and direct-photon production cross
sections per nucleon in $p$Cu to those in $p$Be collisions at 800~GeV/$c$.  
Simple straight line fits to regions with relatively flat distributions 
have been overlaid on the data.
The error bars represent only statistical contributions to
the uncertainties. Systematic uncertainties are indicated by the
shaded region associated with the fit.}
\label{fig:xs_cube_800}
\end{figure}
%%%%%%%%%%%%%%

%%%%%%%%%%%%%%
\begin{figure}
\epsfxsize=\figsize
\epsfbox[\bbcoord]{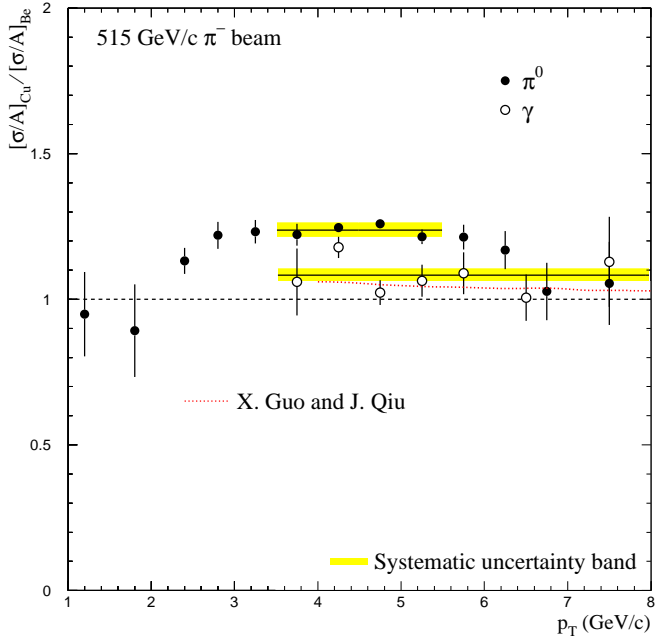}
\caption{
The ratio of inclusive $\pi^0$ and direct-photon production cross
sections per nucleon in $\pi^-$Cu to those in $\pi^-$Be collisions at
515~GeV/$c$. A theoretical prediction for direct-photon production
from X.~Guo and J.~Qiu~\cite{guo} is overlaid on the data (dotted
curve).  Simple straight line fits to regions with relatively flat
distributions have also been overlaid on the data.  The error bars
represent only statistical contributions to the uncertainties.
Systematic uncertainties are indicated by the shaded region associated
with the fit.}
\label{fig:xs_cube_515}
\end{figure}
%%%%%%%%%%%%%%

%%%%%%%%%%%%%%
\begin{figure}
\epsfxsize=\figsize
\epsfbox[\bbcoord]{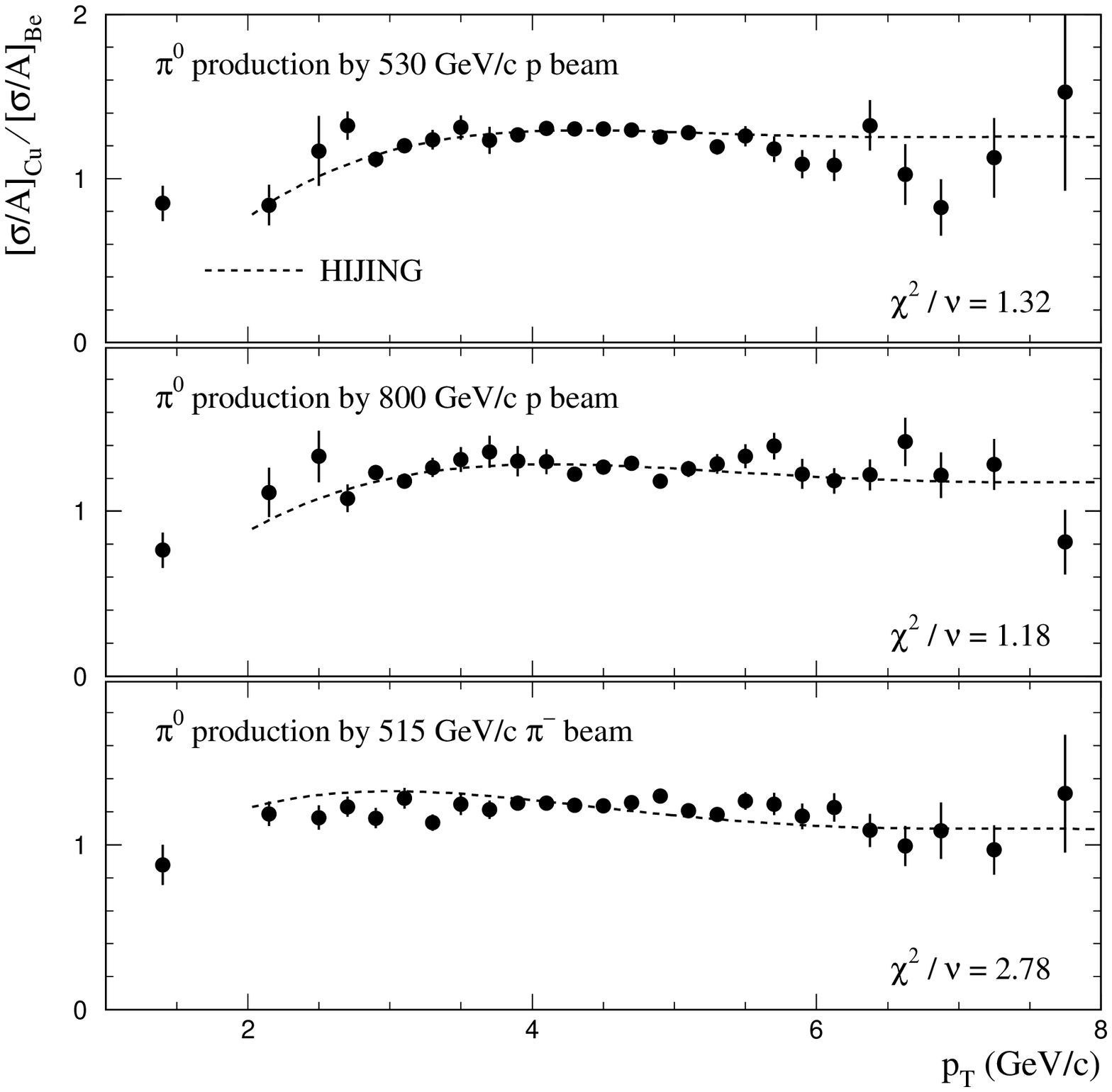}
\caption{
The ratio of inclusive $\pi^0$ production cross sections per nucleon
on Cu to those on Be, compared with predictions from {\sc hijing} scaled
to the data. The error bars represent only statistical contributions
to the uncertainties and $\nu$ represents the number of degrees of freedom
of the fit.
\label{fig:nucl_dep_hijing_pi0}}
\end{figure}

\subsection{Comparisons with {\sc hijing}}

The Cu to Be cross-section ratios are compared with results from the
{\sc hijing} Monte Carlo event generator in
Figs.~\ref{fig:nucl_dep_hijing_pi0} and~\ref{fig:nucl_dep_hijing_g}.
{\sc hijing} is a program designed to simulate particle production in
$pp$, $pA$, and $AA$ collisions~\cite{wang-note}.  It was necessary to
normalize the {\sc hijing} results in Figs.~\ref{fig:nucl_dep_hijing_pi0} and~\ref{fig:nucl_dep_hijing_g} to the data.
However, the shapes of the curves, in the case of the proton beams, are
in good agreement with the data.  
Renewed interest in the amount of nuclear enhancement as a function of
rapidity has been generated by recent BRAHMS
measurements~\cite{brahms-rapidity} and corresponding results from the
other RHIC experiments~\cite{star,phenix,phobos}.  Our high-statistics
cross-section measurements in the central rapidity region can be used
to tune theoretical models developed to describe the RHIC environment.
The rapidity dependence of the Cu to Be cross-section ratios are
compared to expectations from {\sc hijing} in
Figs.~\ref{fig:nucl_dep_rap_530} to \ref{fig:nucl_dep_rap_515}.  {\sc
hijing} does not describe the rapidity dependence of the $\pi^0$ data
for the incident proton beams; the {\sc hijing} results are generally 
peaked towards backward rapidities 
(like the BRAHMS data), whereas our data are relatively independent of rapidity.
{\sc hijing} provides a better description of our direct-photon data.

%%%%%%%%%%%%%%
\begin{figure}
\epsfxsize=\figsize
\epsfbox[\bbcoord]{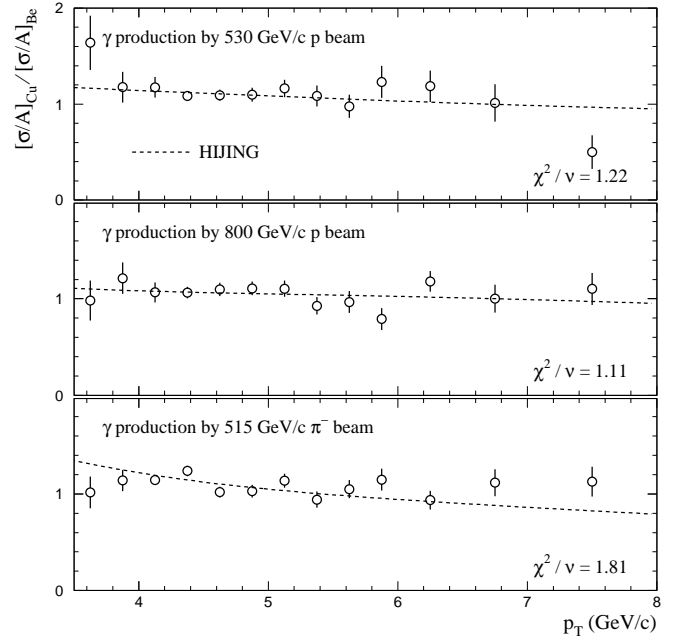}
\caption{
The ratio of inclusive direct-photon production cross sections per nucleon
on Cu to those on Be, compared with predictions
from {\sc hijing} scaled to the data. 
The error bars represent only statistical contributions
to the uncertainties and $\nu$ represents the number of degrees of freedom
of the fit.
\label{fig:nucl_dep_hijing_g}}
\end{figure}

%%%%%%%%%%%%%%%
\begin{figure}
\epsfxsize=\figsize
\epsfbox[\bbcoord]{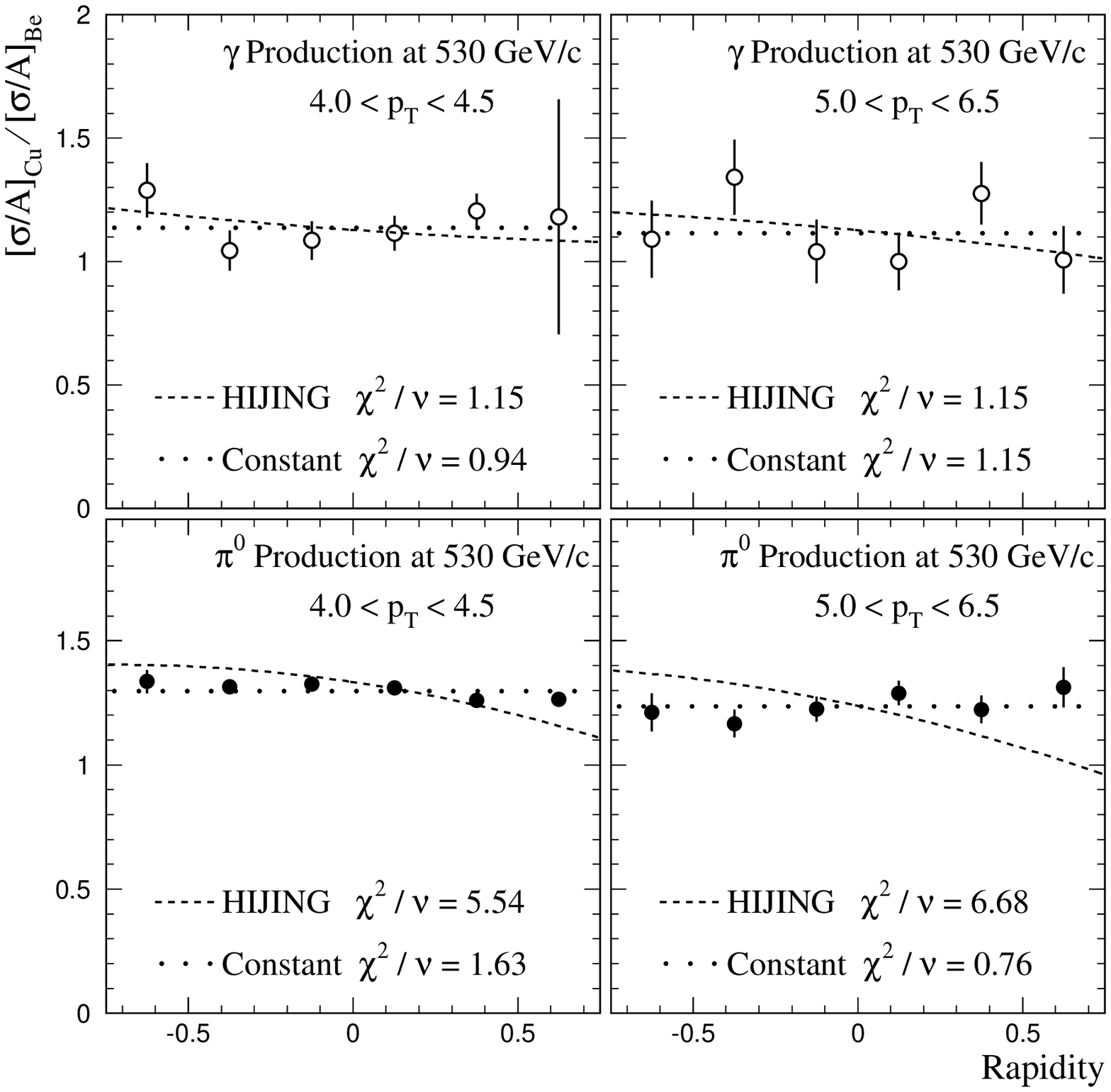}
\caption{
The rapidity dependence of the ratio of inclusive 
cross sections per nucleon on Cu to those on Be 
for direct-photon and $\pi^0$ production in the 530 GeV/$c$
$p$ beam, compared with predictions from {\sc hijing} scaled to the data. 
The error bars represent only statistical contributions to the uncertainties
and $\nu$ represents the number of degrees of freedom of the fit.}
\label{fig:nucl_dep_rap_530}
\end{figure}
%%%%%%%%%%%%%%%
\begin{figure}
\epsfxsize=\figsize
\epsfbox[\bbcoord]{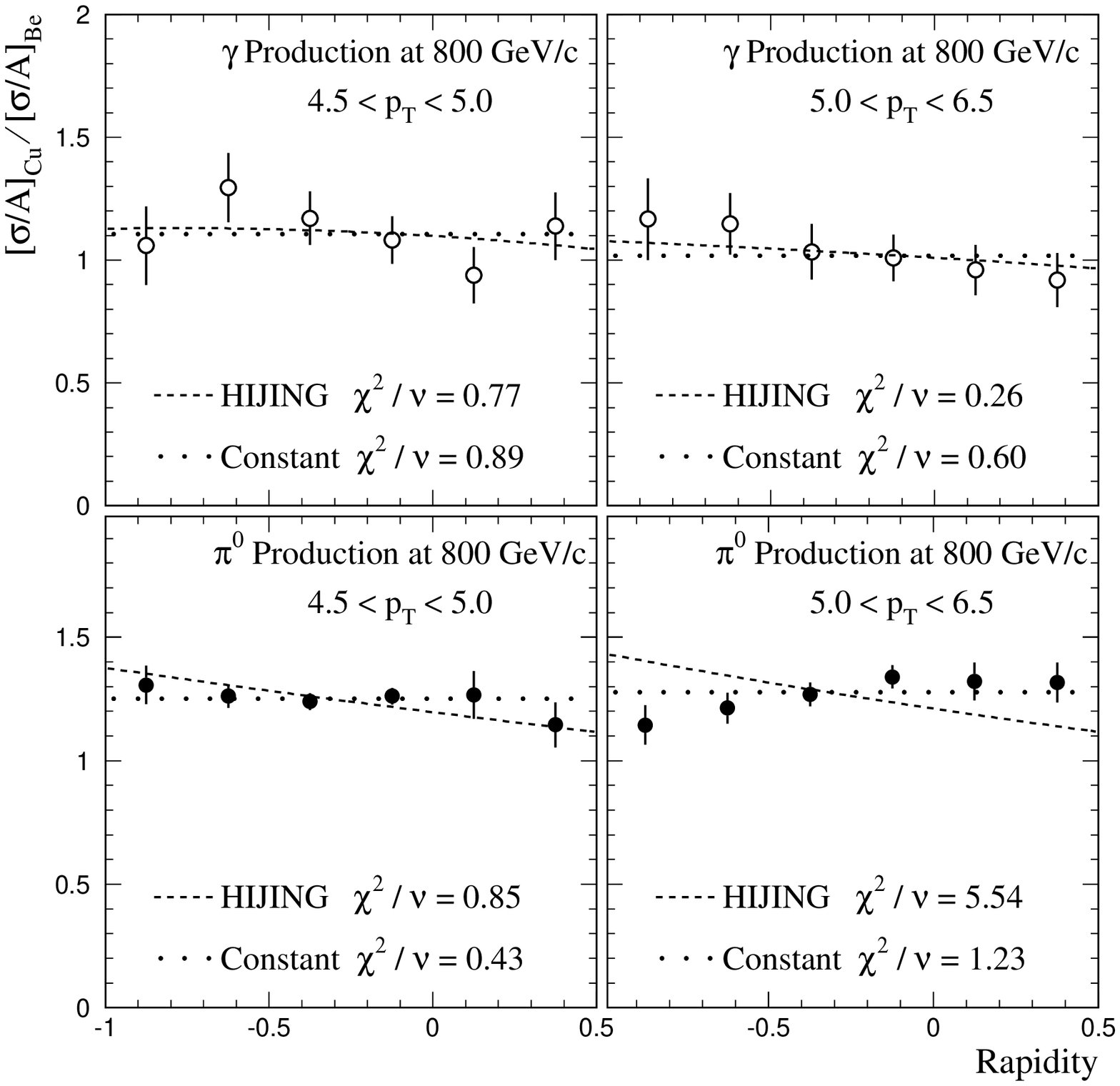}
\caption{
The rapidity dependence of the ratio of inclusive 
cross sections per nucleon on Cu to those on Be for 
direct-photon and $\pi^0$ production in the 800 GeV/$c$ $p$ beam, 
compared with predictions from {\sc hijing} scaled to the data. 
The error bars represent only statistical contributions to the uncertainties
and $\nu$ represents the number of degrees of freedom of the fit.}
\label{fig:nucl_dep_rap_800}
\end{figure}
%%%%%%%%%%%%%%%
\begin{figure}
\epsfxsize=\figsize
\epsfbox[\bbcoord]{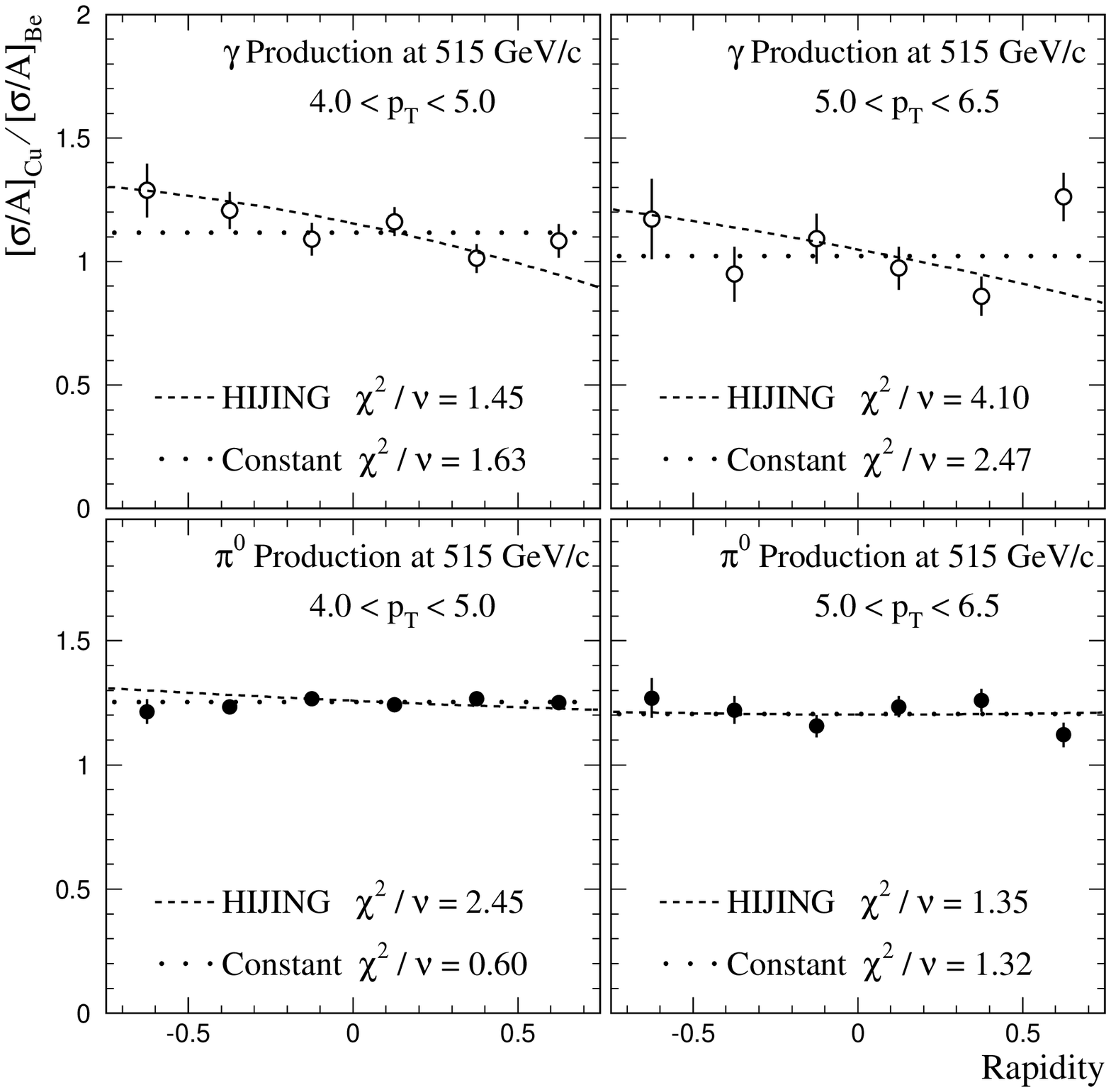}
\caption{
The rapidity dependence of the ratio of inclusive 
cross sections per nucleon on Cu to those on Be 
for direct-photon and $\pi^0$ production in the 515 GeV/$c$
$\pi^-$ beam, compared with predictions from {\sc hijing} scaled
to the data. The error bars
represent only statistical contributions to the uncertainties
and $\nu$ represents the number of degrees of freedom of the fit.}
\label{fig:nucl_dep_rap_515}
\end{figure}
%%%%%%%%%%%%%%%
\section{Conclusions}

We have measured the invariant differential cross section
per nucleon for direct-photon, $\pi^0$, and $\eta$ production from
515~GeV/$c$ $\pi^-$ beam and 800 and 530~GeV/$c$ proton beams incident
on copper as a function of $p_T$ and \ycm.  These data
span the kinematic range $1.0 < p_T \alt 10$~GeV/$c$ and central rapidities.

Ratios of these production cross sections to our previously published
measurements on a beryllium
target~\cite{E706-directphoton,E706-pos-pieta,E706-neg-pieta} show a
strong nuclear enhancement for $\pi^0$ mesons and a smaller, but
significant, enhancement for direct photons.  We compare these
measurements with expectations from a theoretical calculation (for
direct-photon production for incident $\pi^-$ beam) and with the 
results of a Monte Carlo event
generator, {\sc hijing}, dedicated to the simulation of nuclear
effects.  {\sc hijing} yields a good description of the shape of the
$p_T$ dependence of the Cu to Be ratios for both $\pi^0$'s and direct
photons in the incident proton beam data samples. {\sc hijing} also
describes the rapidity dependence of direct photon production in those
samples. However, {\sc hijing} provides a relatively poor description of the
rapidity dependence of $\pi^0$ meson production.

\acknowledgments

We thank the U.~S. Department of Energy, the National Science
Foundation, including its Office of International Programs, and the
Universities Grants Commission of India, for their support of this
research.  The staff and management of Fermilab are thanked for their
efforts in making available the beam and computing facilities that
made this work possible.  We are pleased to acknowledge the
contributions of our colleagues on Fermilab experiment E672. We
acknowledge the contributions of the following colleagues to
the operation and upgrade of the Meson West spectrometer:
W.~Dickerson and E.~Pothier from Northeastern University;
J.~T.~Anderson,
E.~Barsotti~Jr.,
H.~Koecher,
P.~Madsen,
D.~Petravick,
R.~Tokarek, J.~Tweed, D.~Allspach, J.~Urbin, and the cryo crews
from Fermi National Accelerator Laboratory;
T.~Haelen, C.~Benson,
L.~Kuntz, and D.~Ruggiero
from the University of Rochester;
the technical staffs of
Michigan State University and
%
%from University of Delhi;
%
Pennsylvania State University for the construction of the straw tubes
and of the University of Pittsburgh for the silicon detectors. We
thank the following commissioning run collaborators for their
invaluable contributions to the hardware and software infrastructure
of the original Meson West spectrometer: G.~Alverson, G.~Ballocchi,
R.~Benson, D.~Berg, D.~Brown, D.~Carey, T.~Chand, C.~Chandlee,
S.~Easo, W.~Faissler, G.~Glass, I.~Kourbanis, A.~Lanaro,
C.~Nelson~Jr., D.~Orris, B.~Rajaram, K.~Ruddick, A.~Sinanidis, and
G.~Wu.  
We also thank X.-N.~Wang for several helpful discussions.

%\bibliography{paper}

\begin{thebibliography}{39}
\expandafter\ifx\csname natexlab\endcsname\relax\def\natexlab#1{#1}\fi
\expandafter\ifx\csname bibnamefont\endcsname\relax
  \def\bibnamefont#1{#1}\fi
\expandafter\ifx\csname bibfnamefont\endcsname\relax
  \def\bibfnamefont#1{#1}\fi
\expandafter\ifx\csname citenamefont\endcsname\relax
  \def\citenamefont#1{#1}\fi
\expandafter\ifx\csname url\endcsname\relax
  \def\url#1{\texttt{#1}}\fi
\expandafter\ifx\csname urlprefix\endcsname\relax\def\urlprefix{URL }\fi
\providecommand{\bibinfo}[2]{#2}
\providecommand{\eprint}[2][]{\url{#2}}

\bibitem[{\citenamefont{Geist et~al.}(1990)}]{geist}
\bibinfo{author}{\bibfnamefont{W.~M.} \bibnamefont{Geist}}
  \bibnamefont{et~al.}, \bibinfo{journal}{Phys. Rept.}
  \textbf{\bibinfo{volume}{197}}, \bibinfo{pages}{263} (\bibinfo{year}{1990}).

\bibitem[{\citenamefont{McCubbin}(1981)}]{mccubbin}
\bibinfo{author}{\bibfnamefont{N.~A.} \bibnamefont{McCubbin}},
  \bibinfo{journal}{Rep. Prog. Phys.} \textbf{\bibinfo{volume}{44}},
  \bibinfo{pages}{65} (\bibinfo{year}{1981}).

\bibitem[{\citenamefont{Owens}(1987)}]{owens}
\bibinfo{author}{\bibfnamefont{J.~F.} \bibnamefont{Owens}},
  \bibinfo{journal}{Rev. Mod. Phys.} \textbf{\bibinfo{volume}{59}},
  \bibinfo{pages}{465} (\bibinfo{year}{1987}).

\bibitem[{\citenamefont{Ferbel and Molzon}(1984)}]{molzon}
\bibinfo{author}{\bibfnamefont{T.}~\bibnamefont{Ferbel}} \bibnamefont{and}
  \bibinfo{author}{\bibfnamefont{W.~R.} \bibnamefont{Molzon}},
  \bibinfo{journal}{Rev. Mod. Phys.} \textbf{\bibinfo{volume}{56}},
  \bibinfo{pages}{181} (\bibinfo{year}{1984}).

\bibitem[{\citenamefont{Apanasevich et~al.}(1999)}]{kt-dp}
\bibinfo{author}{\bibfnamefont{L.}~\bibnamefont{Apanasevich}}
  \bibnamefont{et~al.}, \bibinfo{journal}{Phys. Rev.}
  \textbf{\bibinfo{volume}{D59}}, \bibinfo{pages}{074007}
  (\bibinfo{year}{1999}).

\bibitem[{\citenamefont{Cronin et~al.}(1975)}]{cronin}
\bibinfo{author}{\bibfnamefont{J.~W.} \bibnamefont{Cronin}}
  \bibnamefont{et~al.}, \bibinfo{journal}{Phys. Rev.}
  \textbf{\bibinfo{volume}{D11}}, \bibinfo{pages}{3105} (\bibinfo{year}{1975}).

\bibitem[{\citenamefont{Antreasyan et~al.}(1979)}]{antreasyan}
\bibinfo{author}{\bibfnamefont{D.}~\bibnamefont{Antreasyan}}
  \bibnamefont{et~al.}, \bibinfo{journal}{Phys. Rev.}
  \textbf{\bibinfo{volume}{D19}}, \bibinfo{pages}{764} (\bibinfo{year}{1979}).

\bibitem[{\citenamefont{Frisch et~al.}(1983)}]{frish}
\bibinfo{author}{\bibfnamefont{H.}~\bibnamefont{Frisch}} \bibnamefont{et~al.},
  \bibinfo{journal}{Phys. Rev.} \textbf{\bibinfo{volume}{D27}},
  \bibinfo{pages}{1001} (\bibinfo{year}{1983}).

\bibitem[{sta()}]{star}
\bibinfo{note}{J. Adams {\it et al.} (STAR), nucl-ex/0501009.}

\bibitem[{phe()}]{phenix}
\bibinfo{note}{K. Adcox {\it et al.} (PHENIX), nucl-ex/0410003.}

\bibitem[{pho()}]{phobos}
\bibinfo{note}{B. B. Back {\it et al.} (PHOBOS), nucl-ex/0410022.}

\bibitem[{bra()}]{brahms}
\bibinfo{note}{I. Arsene {\it et al.} (BRAHMS), nucl-ex/0410020.}

\bibitem[{wan()}]{wang-note}
\bibinfo{note}{X.-N. Wang, Nucl. Phys. {\bf A661}, 609 (1999); Private
  communication with Xin-Nian Wang regarding the use of this program at these
  beam energies.}

\bibitem[{\citenamefont{Apanasevich et~al.}(1998{\natexlab{a}})}]{E706-kt}
\bibinfo{author}{\bibfnamefont{L.}~\bibnamefont{Apanasevich}}
  \bibnamefont{et~al.} (\bibinfo{collaboration}{E706}), \bibinfo{journal}{Phys.
  Rev. Lett.} \textbf{\bibinfo{volume}{81}}, \bibinfo{pages}{2642}
  (\bibinfo{year}{1998}{\natexlab{a}}).

\bibitem[{E70({\natexlab{a}})}]{E706-pos-pieta}
\bibinfo{note}{L. Apanasevich et al. (E706), Phys. Rev. {\bf D68}, 052001
  (2003). A computational error in the correction of measured cross sections
  for contributions from minority particles in secondary beams affected the
  reported cross sections for $\pi^0$ and $\eta$ production by 530~GeV/$c$
  protons. As a result, the cross sections listed in the tables for 530~GeV/$c$
  protons in the referenced publication should be increased by 3.4$\%$, a
  change that falls well within the quoted systematic uncertainties.}

\bibitem[{\citenamefont{Apanasevich
  et~al.}(2004{\natexlab{a}})}]{E706-neg-pieta}
\bibinfo{author}{\bibfnamefont{L.}~\bibnamefont{Apanasevich}}
  \bibnamefont{et~al.} (\bibinfo{collaboration}{E706}), \bibinfo{journal}{Phys.
  Rev.} \textbf{\bibinfo{volume}{D69}}, \bibinfo{pages}{032003}
  (\bibinfo{year}{2004}{\natexlab{a}}).

\bibitem[{\citenamefont{Apanasevich
  et~al.}(2004{\natexlab{b}})}]{E706-directphoton}
\bibinfo{author}{\bibfnamefont{L.}~\bibnamefont{Apanasevich}}
  \bibnamefont{et~al.} (\bibinfo{collaboration}{E706}), \bibinfo{journal}{Phys.
  Rev.} \textbf{\bibinfo{volume}{D70}}, \bibinfo{pages}{092009}
  (\bibinfo{year}{2004}{\natexlab{b}}).

\bibitem[{\citenamefont{Apanasevich et~al.}(1997)}]{E706-charm}
\bibinfo{author}{\bibfnamefont{L.}~\bibnamefont{Apanasevich}}
  \bibnamefont{et~al.} (\bibinfo{collaboration}{E706}), \bibinfo{journal}{Phys.
  Rev.} \textbf{\bibinfo{volume}{D56}}, \bibinfo{pages}{1391}
  (\bibinfo{year}{1997}).

\bibitem[{E70({\natexlab{b}})}]{E706-omega}
\bibinfo{note}{L. Apanasevich {\it et al.} (E706), hep-ex/0004012.}

\bibitem[{\citenamefont{Apanasevich
  et~al.}(1998{\natexlab{b}})}]{E706-calibration}
\bibinfo{author}{\bibfnamefont{L.}~\bibnamefont{Apanasevich}}
  \bibnamefont{et~al.} (\bibinfo{collaboration}{E706}), \bibinfo{journal}{Nucl.
  Inst. \& Meth.} \textbf{\bibinfo{volume}{A417}}, \bibinfo{pages}{50}
  (\bibinfo{year}{1998}{\natexlab{b}}).

\bibitem[{\citenamefont{Allspach et~al.}(1991)}]{h2target}
\bibinfo{author}{\bibfnamefont{D.}~\bibnamefont{Allspach}}
  \bibnamefont{et~al.}, \bibinfo{journal}{Adv. Cryog. Eng.}
  \textbf{\bibinfo{volume}{37}}, \bibinfo{pages}{1495} (\bibinfo{year}{1991}).

\bibitem[{\citenamefont{Carroll et~al.}(1979)}]{carroll}
\bibinfo{author}{\bibfnamefont{A.~S.} \bibnamefont{Carroll}}
  \bibnamefont{et~al.}, \bibinfo{journal}{Phys. Lett.}
  \textbf{\bibinfo{volume}{80B}}, \bibinfo{pages}{319} (\bibinfo{year}{1979}).

\bibitem[{\citenamefont{Baldini et~al.}(1988)}]{Baldini:1988ti}
\bibinfo{author}{\bibfnamefont{A.}~\bibnamefont{Baldini}} \bibnamefont{et~al.},
  in \emph{\bibinfo{booktitle}{Total Cross-Sections for Reactions of High
  Energy Particles}}, edited by
  \bibinfo{editor}{\bibfnamefont{H.}~\bibnamefont{Schopper}}
  (\bibinfo{publisher}{Springer-Verlag}, \bibinfo{address}{New York},
  \bibinfo{year}{1988}), vol.~\bibinfo{volume}{12}.

\bibitem[{\citenamefont{Bromberg et~al.}(1991)}]{E706-STDC}
\bibinfo{author}{\bibfnamefont{C.}~\bibnamefont{Bromberg}} \bibnamefont{et~al.}
  (\bibinfo{collaboration}{E706}), \bibinfo{journal}{Nucl. Inst. \& Meth.}
  \textbf{\bibinfo{volume}{A307}}, \bibinfo{pages}{292} (\bibinfo{year}{1991}).

\bibitem[{\citenamefont{Striley}(1996)}]{striley}
\bibinfo{author}{\bibfnamefont{D.}~\bibnamefont{Striley}}, Ph.D. thesis,
  \bibinfo{school}{University of Missouri at Columbia} (\bibinfo{year}{1996}).

\bibitem[{E70({\natexlab{c}})}]{E706-trigger}
\bibinfo{note}{L.~Sorrell, {\it The E706 Trigger System}, E706 Note 201, 1994
  (unpublished); Ph.D. thesis, Michigan State University, 1995; G.~Osborne,
  Ph.D. thesis, University of Rochester, 1996.}

\bibitem[{BH()}]{BH}
\bibinfo{note}{During the 1990 run, the beam hole definition was implemented
  using a single scintillation counter. An array of four scintillation counters
  was used for the beam hole definition during the 1991--92 run.}

\bibitem[{\citenamefont{Blusk}(1995)}]{blusk}
\bibinfo{author}{\bibfnamefont{S.}~\bibnamefont{Blusk}}, Ph.D. thesis,
  \bibinfo{school}{University of Pittsburgh} (\bibinfo{year}{1995}).

\bibitem[{\citenamefont{Groom et~al.}(2000)}]{pdg}
\bibinfo{author}{\bibfnamefont{D.~E.} \bibnamefont{Groom}} \bibnamefont{et~al.}
  (\bibinfo{collaboration}{Particle Data Group}), \bibinfo{journal}{Eur. Phys.
  J.} \textbf{\bibinfo{volume}{C15}}, \bibinfo{pages}{1}
  (\bibinfo{year}{2000}).

\bibitem[{gea()}]{geant}
\bibinfo{note}{F. Carminati and others, GEANT: Detector Description and
  Simulation Tool, {\rm CERN Program Library Long Writeup W5013}, 1993.}

\bibitem[{\citenamefont{Apanasevich}(2005)}]{apana}
\bibinfo{author}{\bibfnamefont{L.}~\bibnamefont{Apanasevich}}, Ph.D. thesis,
  \bibinfo{school}{Michigan State University} (\bibinfo{year}{2005}).

\bibitem[{\citenamefont{Marchesini et~al.}(1992)}]{herwig56}
\bibinfo{author}{\bibfnamefont{G.}~\bibnamefont{Marchesini}}
  \bibnamefont{et~al.}, \bibinfo{journal}{Comput.~Phys.~Commun.}
  \textbf{\bibinfo{volume}{67}}, \bibinfo{pages}{465} (\bibinfo{year}{1992}),
  \bibinfo{note}{{\sc herwig}~v5.6}.

\bibitem[{\citenamefont{Begel}(1999)}]{begel}
\bibinfo{author}{\bibfnamefont{M.}~\bibnamefont{Begel}}, Ph.D. thesis,
  \bibinfo{school}{University of Rochester} (\bibinfo{year}{1999}).

\bibitem[{\citenamefont{Diakonou et~al.}(1980)}]{etaprimeratio}
\bibinfo{author}{\bibfnamefont{M.}~\bibnamefont{Diakonou}} \bibnamefont{et~al.}
  (\bibinfo{collaboration}{R806}), \bibinfo{journal}{Phys. Lett.}
  \textbf{\bibinfo{volume}{89B}}, \bibinfo{pages}{432} (\bibinfo{year}{1980}).

\bibitem[{hyd()}]{hydrogen-correction}
\bibinfo{note}{Our previously published cross sections on hydrogen were not
  corrected for these misidentified vertices (the corresponding corrections in
  the case of Be were negligible). The applicable correction factors are $1.01$
  for the 1991--92 $\pi^-p$ sample at 515~GeV/c, $1.02$ for the 530~GeV/$c$
  $pp$ sample, and $1.06$ for the 800~GeV/$c$ $pp$ sample. In each case, the
  additional correction falls within our published systematic uncertainty.}

\bibitem[{\citenamefont{Lafferty and Wyatt}(1995)}]{laff}
\bibinfo{author}{\bibfnamefont{G.~D.} \bibnamefont{Lafferty}} \bibnamefont{and}
  \bibinfo{author}{\bibfnamefont{T.~R.} \bibnamefont{Wyatt}},
  \bibinfo{journal}{Nucl. Instrum. Methods Phys. Res.}
  \textbf{\bibinfo{volume}{A355}}, \bibinfo{pages}{541} (\bibinfo{year}{1995}).

\bibitem[{\citenamefont{Wang}(1997)}]{wang-rep}
\bibinfo{author}{\bibfnamefont{X.-N.} \bibnamefont{Wang}},
  \bibinfo{journal}{Phys. Rept.} \textbf{\bibinfo{volume}{280}},
  \bibinfo{pages}{287} (\bibinfo{year}{1997}).

\bibitem[{\citenamefont{Guo and Qiu}(1996)}]{guo}
\bibinfo{author}{\bibfnamefont{X.}~\bibnamefont{Guo}} \bibnamefont{and}
  \bibinfo{author}{\bibfnamefont{J.}~\bibnamefont{Qiu}},
  \bibinfo{journal}{Phys. Rev.} \textbf{\bibinfo{volume}{D53}},
  \bibinfo{pages}{6144} (\bibinfo{year}{1996}).

\bibitem[{\citenamefont{Arsene et~al.}(2004)}]{brahms-rapidity}
\bibinfo{author}{\bibfnamefont{I.}~\bibnamefont{Arsene}} \bibnamefont{et~al.}
  (\bibinfo{collaboration}{BRAHMS}), \bibinfo{journal}{Phys. Rev. Lett.}
  \textbf{\bibinfo{volume}{93}}, \bibinfo{pages}{242303}
  (\bibinfo{year}{2004}).

\end{thebibliography}
%\bibliographystyle{apsrev}

\vfil

\newpage
\clearpage

\onecolumngrid

\appendix*

\section{Tabulated Cross Sections}

In this appendix, we present tables of the measured invariant differential 
cross sections for direct photon, $\pi^0$, and $\eta$ production on Cu targets
as functions of $p_T$. In these tables, the first uncertainty
is statistical and the second is systematic. In the case of the lowest
two $p_T$ bins for the $\pi^0$ measurement, the statistical and 
systematic uncertainties have been combined because of the large 
correlation between them.

\begin{table}[h]
\caption{Invariant differential cross sections $\left( \DIFFXS \right)$ 
per nucleon for direct-photon production in \pCu\ collisions at 800 and
530~GeV/$c$, and \piCu\ collisions at 515~GeV/$c$ as functions of $p_T$.}
\begin{tabular}{r@{ -- }l r@{ }l r@{ }l r@{ }l}
\hline
\hline
\multicolumn{2}{c}{$p_T$}   &\multicolumn{2}{c}{${\mit p}$Cu at 530~GeV/$c$}   &\multicolumn{2}{c}{${\mit p}$Cu at 800~GeV/$c$}   &\multicolumn{2}{c}{${\mit \pi^{-}}$Cu at 515~GeV/$c$} \\
\multicolumn{2}{c}{ }   &\multicolumn{2}{c}{$-0.75\le\ycm\le 0.75$}   &\multicolumn{2}{c}{$-1.0\le\ycm\le 0.5$}   &\multicolumn{2}{c}{$-0.75\le\ycm\le 0.75$} \\
\multicolumn{2}{c}{(GeV/$c$)}    &\multicolumn{2}{c}{$[\rm{nb}$/(GeV/$c)^2]$}    &\multicolumn{2}{c}{$[\rm{nb}$/(GeV/$c)^2]$}    &\multicolumn{2}{c}{$[\rm{nb}$/(GeV/$c)^2]$}  \\
\hline
3.50 & 3.75 & 
 {3.11${\,}\pm$} & {0.45${}\pm{}$0.63} &  {3.01${\,}\pm$} & {0.59${}\pm{}$0.63} &  {1.84${\,}\pm$} & {0.28${}\pm{}$0.34} \\
3.75 & 4.00 & 
 {1.21${\,}\pm$} & {0.15${}\pm{}$0.22} &  {1.87${\,}\pm$} & {0.23${}\pm{}$0.36} &  {1.065${\,}\pm$} & {0.095${}\pm{}$0.18} \\
4.00 & 4.25 & 
 {0.583${\,}\pm$} & {0.048${}\pm{}$0.099} &  {0.989${\,}\pm$} & {0.089${}\pm{}$0.17} &  {0.585${\,}\pm$} & {0.022${}\pm{}$0.093} \\
4.25 & 4.50 & 
 {0.282${\,}\pm$} & {0.011${}\pm{}$0.045} &  {0.523${\,}\pm$} & {0.026${}\pm{}$0.086} &  {0.361${\,}\pm$} & {0.015${}\pm{}$0.053} \\
4.50 & 4.75 & 
 {0.1516${\,}\pm$} & {0.0074${}\pm{}$0.023} &  {0.299${\,}\pm$} & {0.017${}\pm{}$0.046} &  {0.1783${\,}\pm$} & {0.0092${}\pm{}$0.025} \\
4.75 & 5.00 & 
 {0.0865${\,}\pm$} & {0.0052${}\pm{}$0.012} &  {0.182${\,}\pm$} & {0.011${}\pm{}$0.026} &  {0.1124${\,}\pm$} & {0.0066${}\pm{}$0.015} \\
\hline
\multicolumn{2}{c}{ }    &\multicolumn{2}{c}{$[\rm{pb}$/(GeV/$c)^2]$}    &\multicolumn{2}{c}{$[\rm{pb}$/(GeV/$c)^2]$}    &\multicolumn{2}{c}{$[\rm{pb}$/(GeV/$c)^2]$}  \\
\hline
5.00 & 5.25 & 
 {50.2${\,}\pm$} & {3.5${}\pm{}$6.7} &  {110.6${\,}\pm$} & {7.6${}\pm{}$15} &  {81.0${\,}\pm$} & {4.8${}\pm{}$10} \\
5.25 & 5.50 & 
 {27.7${\,}\pm$} & {2.5${}\pm{}$3.6} &  {59.9${\,}\pm$} & {5.5${}\pm{}$7.9} &  {40.4${\,}\pm$} & {3.4${}\pm{}$4.9} \\
5.50 & 5.75 & 
 {15.6${\,}\pm$} & {1.8${}\pm{}$2.0} &  {37.0${\,}\pm$} & {4.1${}\pm{}$4.7} &  {32.3${\,}\pm$} & {2.7${}\pm{}$3.9} \\
5.75 & 6.00 & 
 {11.2${\,}\pm$} & {1.3${}\pm{}$1.4} &  {21.4${\,}\pm$} & {2.9${}\pm{}$2.7} &  {23.1${\,}\pm$} & {2.1${}\pm{}$2.7} \\
6.00 & 6.50 & 
 {4.99${\,}\pm$} & {0.62${}\pm{}$0.61} &  {15.4${\,}\pm$} & {1.3${}\pm{}$1.9} &  {9.74${\,}\pm$} & {0.95${}\pm{}$1.1} \\
6.50 & 7.00 & 
 {1.93${\,}\pm$} & {0.34${}\pm{}$0.24} &  {5.43${\,}\pm$} & {0.72${}\pm{}$0.64} &  {5.30${\,}\pm$} & {0.60${}\pm{}$0.61} \\
7.00 & 8.00 & 
 {0.278${\,}\pm$} & {0.094${}\pm{}$0.035} &  {2.01${\,}\pm$} & {0.27${}\pm{}$0.23} &  {1.84${\,}\pm$} & {0.23${}\pm{}$0.22} \\
8.00 & 10.00 & 
 {0.086${\,}\pm$} & {0.029${}\pm{}$0.012} &  {0.136${\,}\pm$} & {0.055${}\pm{}$0.016} &  {0.197${\,}\pm$} & {0.052${}\pm{}$0.024} \\
10.00 & 12.00 & 
 & {} &  {0.0077${\,}\pm$} & {0.0077${}\pm{}$0.0009} &  {0.0124${\,}\pm$} & {0.0097${}\pm{}$0.0016} \\
\hline
\hline
\end{tabular}
\label{table_gxs_pt_cu}
\end{table}

\begin{table}[h]
\caption{Invariant differential cross sections $\left( \DIFFXS \right)$ 
per nucleon for $\pi^0$ production in \pCu\ collisions at 800 and
530~GeV/$c$, and \piCu\ collisions at 515~GeV/$c$ as functions of $p_T$.}
\begin{tabular}{r@{ -- }l r@{ }l r@{ }l r@{ }l}
\hline
\hline
\multicolumn{2}{c}{$p_T$}   &\multicolumn{2}{c}{${\mit p}$Cu at 530~GeV/$c$}   &\multicolumn{2}{c}{${\mit p}$Cu at 800~GeV/$c$}   &\multicolumn{2}{c}{${\mit \pi^{-}}$Cu at 515~GeV/$c$} \\
\multicolumn{2}{c}{ }   &\multicolumn{2}{c}{$-0.75\le\ycm\le 0.75$}   &\multicolumn{2}{c}{$-1.0\le\ycm\le 0.5$}   &\multicolumn{2}{c}{$-0.75\le\ycm\le 0.75$} \\
\multicolumn{2}{c}{(GeV/$c$)}    &\multicolumn{2}{c}{$[\rm{\mu{b}}$/(GeV/$c)^2]$}    &\multicolumn{2}{c}{$[\rm{\mu{b}}$/(GeV/$c)^2]$}    &\multicolumn{2}{c}{$[\rm{\mu{b}}$/(GeV/$c)^2]$}  \\
\hline
1.00 & 1.40 & 
 {315${\,}\pm$} & {50} &  {293${\,}\pm$} & {67} &  {166${\,}\pm$} & {32} \\
1.40 & 1.80 & 
 {43.2${\,}\pm$} & {9.9} &  {79${\,}\pm$} & {16} &  {23.9${\,}\pm$} & {6.2} \\
1.80 & 2.20 & 
 {7.3${\,}\pm$} & {2.6${}\pm{}$0.8} &  {18.5${\,}\pm$} & {4.1${}\pm{}$2.0} &  {8.72${\,}\pm$} & {0.88${}\pm{}$1.0} \\
2.20 & 2.40 & 
 {2.64${\,}\pm$} & {0.19${}\pm{}$0.28} &  {3.77${\,}\pm$} & {0.40${}\pm{}$0.42} &  {2.015${\,}\pm$} & {0.094${}\pm{}$0.24} \\
2.40 & 2.60 & 
 {1.38${\,}\pm$} & {0.23${}\pm{}$0.15} &  {2.00${\,}\pm$} & {0.20${}\pm{}$0.22} &  {0.973${\,}\pm$} & {0.057${}\pm{}$0.11} \\
2.60 & 2.80 & 
 {0.663${\,}\pm$} & {0.041${}\pm{}$0.070} &  {0.797${\,}\pm$} & {0.053${}\pm{}$0.088} &  {0.505${\,}\pm$} & {0.023${}\pm{}$0.058} \\
2.80 & 3.00 & 
 {0.2723${\,}\pm$} & {0.0093${}\pm{}$0.029} &  {0.476${\,}\pm$} & {0.014${}\pm{}$0.052} &  {0.244${\,}\pm$} & {0.012${}\pm{}$0.028} \\
3.00 & 3.20 & 
 {0.1453${\,}\pm$} & {0.0048${}\pm{}$0.016} &  {0.2287${\,}\pm$} & {0.0082${}\pm{}$0.025} &  {0.1417${\,}\pm$} & {0.0062${}\pm{}$0.016} \\
\hline
\multicolumn{2}{c}{ }    &\multicolumn{2}{c}{$[\rm{nb}$/(GeV/$c)^2]$}    &\multicolumn{2}{c}{$[\rm{nb}$/(GeV/$c)^2]$}    &\multicolumn{2}{c}{$[\rm{nb}$/(GeV/$c)^2]$}  \\
\hline
3.20 & 3.40 & 
 {77.5${\,}\pm$} & {3.2${}\pm{}$8.2} &  {132.5${\,}\pm$} & {5.3${}\pm{}$15} &  {69.2${\,}\pm$} & {2.7${}\pm{}$7.7} \\
3.40 & 3.60 & 
 {42.0${\,}\pm$} & {2.1${}\pm{}$4.5} &  {70.1${\,}\pm$} & {3.5${}\pm{}$7.7} &  {39.1${\,}\pm$} & {2.0${}\pm{}$4.3} \\
3.60 & 3.80 & 
 {20.0${\,}\pm$} & {1.2${}\pm{}$2.1} &  {41.2${\,}\pm$} & {2.6${}\pm{}$4.5} &  {21.20${\,}\pm$} & {0.92${}\pm{}$2.4} \\
3.80 & 4.00 & 
 {11.53${\,}\pm$} & {0.34${}\pm{}$1.2} &  {23.1${\,}\pm$} & {1.4${}\pm{}$2.5} &  {12.27${\,}\pm$} & {0.37${}\pm{}$1.4} \\
4.00 & 4.20 & 
 {6.411${\,}\pm$} & {0.067${}\pm{}$0.69} &  {12.74${\,}\pm$} & {0.68${}\pm{}$1.4} &  {6.947${\,}\pm$} & {0.075${}\pm{}$0.77} \\
4.20 & 4.40 & 
 {3.579${\,}\pm$} & {0.043${}\pm{}$0.39} &  {6.89${\,}\pm$} & {0.15${}\pm{}$0.76} &  {4.021${\,}\pm$} & {0.054${}\pm{}$0.44} \\
4.40 & 4.60 & 
 {2.002${\,}\pm$} & {0.030${}\pm{}$0.22} &  {4.15${\,}\pm$} & {0.12${}\pm{}$0.46} &  {2.389${\,}\pm$} & {0.035${}\pm{}$0.26} \\
4.60 & 4.80 & 
 {1.146${\,}\pm$} & {0.020${}\pm{}$0.13} &  {2.528${\,}\pm$} & {0.075${}\pm{}$0.28} &  {1.438${\,}\pm$} & {0.024${}\pm{}$0.16} \\
4.80 & 5.00 & 
 {0.639${\,}\pm$} & {0.014${}\pm{}$0.070} &  {1.426${\,}\pm$} & {0.041${}\pm{}$0.16} &  {0.892${\,}\pm$} & {0.018${}\pm{}$0.098} \\
\hline
\multicolumn{2}{c}{ }    &\multicolumn{2}{c}{$[\rm{pb}$/(GeV/$c)^2]$}    &\multicolumn{2}{c}{$[\rm{pb}$/(GeV/$c)^2]$}    &\multicolumn{2}{c}{$[\rm{pb}$/(GeV/$c)^2]$}  \\
\hline
5.00 & 5.20 & 
 {392${\,}\pm$} & {11${}\pm{}$43} &  {931${\,}\pm$} & {31${}\pm{}$110} &  {514${\,}\pm$} & {13${}\pm{}$57} \\
5.20 & 5.40 & 
 {215.8${\,}\pm$} & {7.6${}\pm{}$24} &  {582${\,}\pm$} & {23${}\pm{}$66} &  {320${\,}\pm$} & {11${}\pm{}$35} \\
5.40 & 5.60 & 
 {134.3${\,}\pm$} & {5.9${}\pm{}$15} &  {374${\,}\pm$} & {18${}\pm{}$42} &  {219.2${\,}\pm$} & {8.3${}\pm{}$24} \\
5.60 & 5.80 & 
 {72.2${\,}\pm$} & {4.3${}\pm{}$8.1} &  {258${\,}\pm$} & {14${}\pm{}$29} &  {130.9${\,}\pm$} & {6.7${}\pm{}$15} \\
5.80 & 6.00 & 
 {44.4${\,}\pm$} & {3.2${}\pm{}$5.0} &  {143.6${\,}\pm$} & {9.3${}\pm{}$16} &  {79.7${\,}\pm$} & {4.7${}\pm{}$8.9} \\
6.00 & 6.25 & 
 {25.2${\,}\pm$} & {2.0${}\pm{}$2.9} &  {88.7${\,}\pm$} & {5.2${}\pm{}$10} &  {51.2${\,}\pm$} & {3.3${}\pm{}$5.8} \\
6.25 & 6.50 & 
 {16.6${\,}\pm$} & {1.7${}\pm{}$1.9} &  {53.9${\,}\pm$} & {3.7${}\pm{}$6.3} &  {27.8${\,}\pm$} & {2.4${}\pm{}$3.2} \\
6.50 & 6.75 & 
 {6.3${\,}\pm$} & {1.0${}\pm{}$0.7} &  {33.8${\,}\pm$} & {3.1${}\pm{}$3.9} &  {15.3${\,}\pm$} & {1.7${}\pm{}$1.7} \\
6.75 & 7.00 & 
 {3.71${\,}\pm$} & {0.71${}\pm{}$0.44} &  {19.1${\,}\pm$} & {1.9${}\pm{}$2.2} &  {9.5${\,}\pm$} & {1.4${}\pm{}$1.1} \\
7.00 & 7.50 & 
 {1.96${\,}\pm$} & {0.38${}\pm{}$0.23} &  {9.5${\,}\pm$} & {1.0${}\pm{}$1.1} &  {4.39${\,}\pm$} & {0.63${}\pm{}$0.51} \\
7.50 & 8.00 & 
 {0.54${\,}\pm$} & {0.18${}\pm{}$0.07} &  {2.10${\,}\pm$} & {0.46${}\pm{}$0.25} &  {1.73${\,}\pm$} & {0.44${}\pm{}$0.21} \\
8.00 & 9.00 & 
 {0.149${\,}\pm$} & {0.061${}\pm{}$0.019} &  {0.95${\,}\pm$} & {0.26${}\pm{}$0.12} &  {0.43${\,}\pm$} & {0.17${}\pm{}$0.05} \\
9.00 & 10.00 & 
 & {} &  {} & {} &  {0.067${\,}\pm$} & {0.047${}\pm{}$0.009} \\
\hline
\hline
\end{tabular}
\label{table_pi0xs_pt_cu}
\end{table}

\begin{table}[h]
\caption{Invariant differential cross sections $\left( \DIFFXS \right)$ 
per nucleon for $\eta$ production in \pCu\ collisions at 800 and
530~GeV/$c$, and \piCu\ collisions at 515~GeV/$c$ as functions of $p_T$.}
\begin{tabular}{r@{ -- }l r@{ }l r@{ }l r@{ }l}
\hline
\hline
\multicolumn{2}{c}{$p_T$}   &\multicolumn{2}{c}{${\mit p}$Cu at 530~GeV/$c$}   &\multicolumn{2}{c}{${\mit p}$Cu at 800~GeV/$c$}   &\multicolumn{2}{c}{${\mit \pi^{-}}$Cu at 515~GeV/$c$} \\
\multicolumn{2}{c}{ }   &\multicolumn{2}{c}{$-0.75\le\ycm\le 0.75$}   &\multicolumn{2}{c}{$-1.0\le\ycm\le 0.5$}   &\multicolumn{2}{c}{$-0.75\le\ycm\le 0.75$} \\
\multicolumn{2}{c}{(GeV/$c$)}    &\multicolumn{2}{c}{$[\rm{nb}$/(GeV/$c)^2]$}    &\multicolumn{2}{c}{$[\rm{nb}$/(GeV/$c)^2]$}    &\multicolumn{2}{c}{$[\rm{nb}$/(GeV/$c)^2]$}  \\
\hline
3.00 & 3.50 & 
 {53${\,}\pm$} & {12${}\pm{}$6.0} &  {82${\,}\pm$} & {25${}\pm{}$10} &  {44${\,}\pm$} & {12${}\pm{}$6.0} \\
3.50 & 4.00 & 
 {7.9${\,}\pm$} & {2.2${}\pm{}$0.9} &  {9.9${\,}\pm$} & {5.1${}\pm{}$1.2} &  {10.6${\,}\pm$} & {1.6${}\pm{}$1.3} \\
4.00 & 4.50 & 
 {1.68${\,}\pm$} & {0.17${}\pm{}$0.19} &  {4.78${\,}\pm$} & {0.47${}\pm{}$0.56} &  {2.21${\,}\pm$} & {0.17${}\pm{}$0.26} \\
4.50 & 5.00 & 
 {0.434${\,}\pm$} & {0.043${}\pm{}$0.050} &  {1.06${\,}\pm$} & {0.12${}\pm{}$0.12} &  {0.730${\,}\pm$} & {0.049${}\pm{}$0.084} \\
\hline
\multicolumn{2}{c}{ }    &\multicolumn{2}{c}{$[\rm{pb}$/(GeV/$c)^2]$}    &\multicolumn{2}{c}{$[\rm{pb}$/(GeV/$c)^2]$}    &\multicolumn{2}{c}{$[\rm{pb}$/(GeV/$c)^2]$}  \\
\hline
5.00 & 5.50 & 
 {120${\,}\pm$} & {13${}\pm{}$14} &  {274${\,}\pm$} & {37${}\pm{}$32} &  {203${\,}\pm$} & {17${}\pm{}$24} \\
5.50 & 6.00 & 
 {41.0${\,}\pm$} & {6.6${}\pm{}$4.9} &  {123${\,}\pm$} & {23${}\pm{}$15} &  {56.6${\,}\pm$} & {8.0${}\pm{}$6.6} \\
6.00 & 7.00 & 
 {7.5${\,}\pm$} & {1.4${}\pm{}$0.9} &  {15.8${\,}\pm$} & {4.7${}\pm{}$1.9} &  {18.2${\,}\pm$} & {2.3${}\pm{}$2.2} \\
7.00 & 8.00 & 
 {1.01${\,}\pm$} & {0.35${}\pm{}$0.13} &  {3.0${\,}\pm$} & {1.5${}\pm{}$0.4} &  {2.66${\,}\pm$} & {0.76${}\pm{}$0.33} \\
8.00 & 10.00 & 
 {0.087${\,}\pm$} & {0.084${}\pm{}$0.012} &  {0.15${\,}\pm$} & {0.41${}\pm{}$0.02} &  {0.29${\,}\pm$} & {0.16${}\pm{}$0.04} \\
\hline
\hline
\end{tabular}
\label{table_etaxs_pt_cu}
\end{table}

%\vfil
\end{document}